\documentclass[12pt]{article}

\usepackage[margin=1in]{geometry}
\usepackage[T1]{fontenc}
\usepackage{amsmath,amssymb,amsthm}
\usepackage{array}
\usepackage{booktabs}
\usepackage{caption}
\usepackage{enumitem}
\usepackage{float}
\usepackage{epigraph}
\usepackage{graphicx}
\usepackage{hyperref} 
\hypersetup{
    colorlinks=true,
    linkcolor=blue,
    urlcolor=blue,
    citecolor=blue
}
\usepackage{natbib}
\usepackage{microtype}
\usepackage{natbib}
\usepackage{setspace}
\usepackage{subcaption}
\usepackage{tikz}
\usepackage{xcolor}
\usetikzlibrary{arrows.meta}

\newtheorem{definition}{Definition}

\newtheorem{hypothesis}{Hypothesis}

\newtheorem{proposition}{Proposition}
\newtheorem{result}{Result}

\setlist{nosep}

\graphicspath{{figures/}}

\title{Algorithm-Driven Information Similarity and Collective Action: An Experimental Study\thanks{The experiment was pre-registered on the AEA registry, number AEARCTR-0018776. The experimental protocol was approved by the IRB at GfeW, number DyG533H5.}}
\author{Manshu Khanna\thanks{Peking University HSBC Business School, Shenzhen 518055, China. Email: manshu@phbs.pku.edu.cn.} \and Bozhang Xia\thanks{Peking University HSBC Business School, Shenzhen 518055, China. Email: bozhang@stu.pku.edu.cn.}}
\date{\today}

\begin{document}

\maketitle

\begin{abstract}

We study how the similarity of individuals' information shapes collective action. When people draw on a common source of information, such as social media, each becomes more confident about what others have seen and will do. This can help them coordinate, but it can also tempt them to free-ride. We show that which force prevails depends on how demanding the collective goal is. In a content-moderation experiment, subjects decide whether to pay a cost to report harmful content, which is removed only if enough reports are received. We vary the similarity of group members' information, holding fixed what each learns on her own, and independently vary the removal threshold. More similar information impedes reporting when few reports suffice and facilitates it when many are required, lowering reporting by 17 percentage points under an easy threshold and raising it by 34 points under a demanding one. This confirms the central comparative static of the theory of information similarity \citep{basakDebKuvalekarCollective}. Elicited beliefs trace the reversal to perceived pivotality and document systematic miscalibration of it. Subjects overestimate pivotality across all regimes, and their beliefs respond to similarity in line with actual pivotality only at intermediate thresholds: easy thresholds produce unrecognized pivotality, and near-unanimous thresholds produce illusory pivotality. The two response-miscalibrated patterns coincide with welfare losses; only under aligned pivotality does greater participation translate into greater collective success and higher welfare.

\end{abstract}


\bigskip

\pagebreak
\onehalfspacing

\noindent\begin{minipage}[t]{0.45\textwidth}
\raggedright
\itshape
``[A]lgorithms proactively amplified and promoted content on the Facebook platform which incited violence, hatred, and discrimination against the Rohingya.''\\[4pt]
\upshape--- Amnesty International \citeyearpar{amnesty2022social}
\end{minipage}\hfill
\begin{minipage}[t]{0.45\textwidth}
\raggedleft
\itshape
``Young people are just on Instagram. This is a
potential segment of citizens that we all need
to connect with.''\\[4pt]
\upshape--- Prime Minister Narendra Modi,\\ to his Cabinet, July 2026
\end{minipage}

\vspace{2em}
\section{Introduction}

In June 2026, a satirical Instagram page called the Cockroach Janta Party (CJP)---created after India's Chief Justice likened unemployed youth to ``cockroaches''---gained millions of followers within days and, within weeks, drew thousands of protesters to Jantar Mantar in New Delhi demanding examination reform and a minister's resignation.\footnote{``The Rise of the Cockroach Janta Party: From Instagram Page to Youth Movement,'' \emph{Asia Media Centre}, June 9, 2026, \url{https://www.asiamediacentre.org.nz/from-screens-to-streets-the-psychology-behind-indias-latest-youth-movement}.} The movement's engine was not its follower base but the platform's recommendation algorithm: 82 percent of engagement with the page's content came from users who did not follow it, as Instagram pushed the same posts onto millions of screens.\footnote{``The NEET Protests Reveal How Creators, Algorithms and Public Sentiment Intersect,'' \emph{BuzzInContent}, July 24, 2026, \url{https://www.buzzincontent.com/insight/the-neet-protests-reveal-how-creators-algorithms-and-public-sentiment-intersect-12193437}.} And because Instagram's users skew overwhelmingly young---most older Indians are absent from the platform---the algorithm did more than spread information; it made an entire generation's information \emph{similar}. The fact was not lost on the Prime Minister, who addressed the protesters in a video posted to the platform and instructed his cabinet ministers to join Instagram and engage with the youth through reels, remarking that ``young people are just on Instagram.''\footnote{``PM Narendra Modi To Ministers: `Engage With Youth On Instagram, Post Reels','' \emph{NDTV}, July 24, 2026, \url{https://www.ndtv.com/india-news/pm-narendra-modi-to-ministers-engage-with-youth-on-instagram-post-reels-11817156}.} Did this algorithmic alignment of information cause the protests to swell---or would the movement have grown larger still had each young Indian been less certain of what everyone else had already seen?

Two influential studies of political mobilization give opposite answers. Examining the diffusion of Russia's dominant social-media platform, \citet{enikolopovMakarinPetrova2020} find that cities whose residents drew on a common information source produced larger protests. Examining protests in Hong Kong, \citet{cantoniEtAl2019} find that potential protesters who learned that others intended to turn out became less likely to turn out themselves. In both settings, people's information about the world, and about one another, became more aligned. In the first, alignment mobilized; in the second, it demobilized. 

These findings are not contradictory. Information shared across people does two things at once. It tells each person something about the state of the world, and it tells her something about what others have probably seen and will probably do. The second effect cuts both ways. Believing that others observed the same reason to act can make people coordinate more readily. The same belief can also convince them that enough others will act without them, so that their own costly action is wasted. Whether shared information mobilizes or demobilizes, therefore, turns on a feature of the institution that has nothing to do with the information itself: how individual actions combine into collective success. The institution enters twice. Through a \emph{participation channel}, it sets whether more similar information encourages or discourages costly participation; through a \emph{conversion channel}, it separately governs whether that participation is turned into collective success. The two need not move together: the information structure that most mobilizes participation can be the one under which the institution least converts it into outcomes.  

This logic has a precise formulation. \citet{basakDebKuvalekarGames} compare information structures while holding fixed each agent's marginal information, varying only what one person's signal implies about the signals of others. \citet{basakDebKuvalekarCollective} bring this comparison to collective action and show that more similar information raises participation when the goal is demanding and lowers participation when the goal is easy. The distinction is not whether people grow more optimistic about the state of the world. It is whether others' expected participation makes one's own action pivotal or redundant.

Establishing this mechanism outside the laboratory is difficult because the forces that make information more similar rarely move similarity alone. A recommendation system that steers users toward the same content also changes what that content is: collaborative filters lead similar consumers to explore the same kinds of products and shift attention toward popular items, while major news platforms homogenize top search results across users \citep{leeHosanagar2019,nechushtaiZamithLewis2024}; studies of Facebook likewise show that algorithmic ranking reduces exposure to cross-cutting news, making users less likely to encounter posts from counter-attitudinal outlets \citep{bakshyMessingAdamic2015,levy2021}. A viral message that coordinates citizens also changes their beliefs about the underlying grievance: communication technology conveys information about both dissatisfaction with the regime and protest logistics \citep{little2016}; public information updates beliefs about government performance and higher-order beliefs about other citizens, thereby facilitating protest \citep{hollyerRosendorffVreeland2015}; mobile-phone access makes individuals more responsive both to economic conditions through enhanced information and to others' participation through enhanced coordination \citep{manacordaTesei2020}. Corporate governance creates the same empirical entanglement: when directors observe a common public signal in addition to their private assessments of CEO quality, their information becomes more similar, but the public signal also changes what each director learns about the CEO \citep{chemmanurFedaseyeu2018}; proxy advisors' recommendations can substantially shift support in say-on-pay votes \citep{malenkoShen2016}, while reliance on a common recommendation can crowd out investors' independent research and generate excessive conformity \citep{malenkoMalenko2019}. Isolating the strategic channel requires making \emph{information more similar} while holding fixed what each person learns about the world, and varying the difficulty of collective success independently of the information structure. We do both.  

But what does it mean to make information more similar? \citet{basakDebKuvalekarGames} introduce an ordering of information structures, \emph{concentration along diagonal} (CAD). Two structures are held to the same marginals---each agent, taken alone, learns exactly as much about the state under one as under the other---and they are ranked only by what one agent's signal reveals about another's. One structure is more similar than a second when, conditional on observing any given signal, an agent regards it as more likely that others observed the same signal and less likely that they observed a different one. The ordering thus changes the strategic content of information while leaving its fundamental content untouched: it varies what each person infers
about others without varying what she infers about the state of the world. This separation makes comparative statics interpretable, since it isolates the channel through which similarity affects coordination and free-riding from changes in how well-informed people are.

We study a content-moderation game with 576 participants. In twelve-person groups, subjects decide whether to pay a cost to report potentially harmful content; the content is removed, and every member benefits, only if reports reach a threshold. Before deciding, each subject sees an algorithmic report. With probability $\rho$ the algorithm conducts a single evaluation and sends the same report to all twelve members; with probability $1-\rho$ it evaluates each subject independently. Raising $\rho$ makes subjects' information more similar in the CAD order but leaves the accuracy of any subject's own report unchanged---the experimental counterpart of the theoretical comparative static. We cross this with the institution that converts reports into removal: the threshold is low (one to four reports), medium (five to eight), or high (nine to twelve). When the threshold is high, others' reports are strategic complements, and a report matters only if many others report as well. When the threshold is low, they are strategic substitutes, and a few other reports already suffice.  

Our central result is a reversal. We focus on subjects who receive a warning, so that the state is effectively known and $\rho$ shifts only beliefs about others. Among these subjects, moving from idiosyncratic to common information lowers reporting by about $17$ percentage points when the threshold is low, but raises it by about $34$ percentage points when the threshold is high. The two slopes have opposite signs and differ by roughly $51$ percentage points: the same change in the information structure mobilizes or demobilizes depending on the institution alone. The exact threshold varies across rounds within each session. We pool observations from all sessions and estimate the similarity slope as a linear function of the exact threshold over the full range. The similarity slope reverses sign at $T^{*}\approx4.5$ in our experimental environment, just over a third of the group.\footnote{In the finite binary-signal setting, \citet{basakDebKuvalekarCollective} predict a single sign reversal as collective success becomes more demanding: more similar information discourages participation when only a few actions are required and encourages it when many actions are required. The precise location of this reversal depends on the distribution of others' participation and the reporting rate sustained in equilibrium. Pooling across sessions, whose regimes between them span the full range of exact thresholds, allows us to locate it in this setting; see Section~\ref{sec:theory}.}

Elicited beliefs trace the reversal to perceived pivotality. More similar information raises the expected number of other reporters in every regime, the mechanical consequence of correlated signals. But an increase in expected participation does not imply the pivotality of one's own costly action. When we ask subjects for the probability that exactly $T-1$ others report---the event in which their own report is pivotal---the response reverses across regimes. Similarity raises perceived pivotality where the threshold is high and lowers it where the threshold is low. This belief about one's own pivotality tracks reporting closely and partially mediates the reversal: adding it to the reporting regression attenuates the High--Low difference in the similarity slope by about one quarter. The contrasting pattern in expected participation is inconsistent with simple imitation. Pivotality calibration has two distinct dimensions. The first is calibration in levels: subjects systematically overestimate how often their own report is pivotal in every threshold regime. The second is response alignment: whether perceived and actual pivotality move in the same direction when information similarity changes. They do so only in the Medium regime. In the Low regime, similarity lowers perceived pivotality while raising actual pivotality, a pattern we call \emph{unrecognized pivotality}. In the Medium regime, it raises both, producing \emph{aligned pivotality}. In the High regime, it raises perceived pivotality without measurably changing actual pivotality, producing \emph{illusory pivotality}. Unrecognized and illusory pivotality are two forms of response miscalibration. These labels describe responses to information similarity, not calibration in levels: subjects overestimate pivotality in all three regimes, including the aligned Medium regime.

These alignment patterns map onto collective outcomes. Under unrecognized pivotality in the Low regime, similarity lowers perceived pivotality and reporting even as actual pivotality becomes more frequent, leaving more groups short of a reachable threshold and reducing welfare. Under aligned pivotality in the Medium regime, perceived and actual pivotality both rise, and additional reports clear reachable thresholds, increasing removal and welfare. Under illusory pivotality in the High regime, perceived pivotality and reporting rise without a corresponding increase in actual pivotality or removal; the additional reports are largely wasted, and welfare falls. Pivotality alignment describes whether perceived influence tracks actual opportunities for consequential action. The conversion channel determines whether the resulting reports are sufficient to reach the threshold. Together, they explain why the same increase in participation can be socially productive in one regime and wasteful in another. The sign of the welfare effect is therefore not set by the direction of the reporting response alone: more similar information lowers welfare both where it discourages reporting, so a reachable threshold slips out of reach, and where it most encourages reporting but a near-unanimous threshold remains beyond reach.

We also show that controlling for incentivized measures of risk, altruism, and trust leaves the crossover essentially unchanged, and that a one-parameter logit QRE reproduces its sign change and approximate location. The QRE falls short in the High regime's illusory-pivotality case, where subjects most overestimate pivotality; an elicited-belief logit recovers the observed reporting level and positive similarity response.

We make three contributions. First, we provide the first direct experimental test of how information similarity shapes collective action. Similarity is a property of the joint information structure---what one person's signal reveals about the signals others receive---and is rarely varied in isolation in field settings. Greater social-media penetration raises protest activity, while mobile-phone coverage does so during economic downturns \citep{enikolopovMakarinPetrova2020,manacordaTesei2020}. However, such variation can change several margins at once: the content, reach, and salience of information, communication costs, network composition, and exposure to common sources. \citet{enikolopovMakarinPetrova2020} carefully distinguish an information channel from a collective-action channel, but provide indirect evidence on the latter from outcome patterns rather than varying the dependence of individuals' information while holding each person's own information fixed. Experimental evidence goes further. \citet{cantoniEtAl2019} cleanly identify how beliefs about others' turnout affect one's own turnout. However, they vary the expected level of participation rather than information similarity, and do so within a single institution.

The open question is therefore not whether communication or information about others matters, but whether more similar information changes participation when what each person learns about the state is held fixed, and whether its effect depends on the institution that converts individual actions into collective success. This is where the theory of information similarity earns its empirical keep. \citet{basakDebKuvalekarGames} and \citet{basakDebKuvalekarCollective} make the loosely invoked coordination channel precise: they pin down both what must be held fixed to separate similarity from better individual information and which institutional feature determines the sign of its behavioral effect. The theory therefore offers a unified interpretation of evidence that otherwise looks contradictory: the same strategic channel---what information implies about others' actions---can generate mobilization or free-riding under different institutions. Our design supplies the counterfactual unavailable in existing field evidence: holding each subject's marginal signal distribution fixed shuts down the fundamental-information channel, random assignment of information structures removes selection into common sources, random rematching eliminates persistent networks, and independent variation in the threshold separates similarity from the rule for collective success. The reporting reversal directly tests the theory's central comparative static. More broadly, the design demonstrates the theory's empirical payoff: it provides evidence consistent with a strategic channel that existing reduced-form estimates do not separately identify and provides a common interpretation of effects that otherwise look context-specific or contradictory.

Second, we distinguish calibration in levels from alignment in responses. Subjects overestimate pivotality in all three regimes, but changes in information similarity move perceived and actual pivotality together only under intermediate thresholds. This distinction yields three empirically distinct patterns---unrecognized, aligned, and illusory pivotality---and shifts attention from whether people overestimate their influence to whether the information structure moves perceived influence in line with actual pivotal opportunities.

Third, we show that these response patterns map systematically onto collective outcomes in our experiment. Welfare rises under aligned pivotality but falls under both forms of response miscalibration, although through different margins: unrecognized pivotality produces forgone collective success, whereas illusory pivotality produces costly but ineffective participation.

\paragraph{Related Literature.}

Our paper is most directly related to the literature on collective action. Collective action requires enough individuals to take costly actions to achieve a shared goal, while the benefits of success accrue broadly, creating a classic free-rider problem \citep{olson1965,tullock1971}. Since success depends on how many others act, information matters not only for what it reveals about the state of the world but for what it implies about others' behavior, which is what makes an agent's own action pivotal \citep{palfreyRosenthal1984,palfreyRosenthal1985,myerson1998}. This is exactly why information about others cuts both ways. On one side, it enables \emph{coordination}: learning that others are likely to act can make one's own participation worthwhile. This channel is emphasized by theories of protest and regime change \citep{edmond2013,battaglini2017} and by field evidence that mobile phones and social media raise turnout \citep{manacordaTesei2020,enikolopovMakarinPetrova2020}. On the other, it enables \emph{free-riding}: the same expectation can make one's costly action redundant. This force is pronounced in models that embed free-riding into participation \citep{dziudaGitmezShadmehr2021,myattWallace2008,shadmehr2021} and in experimental evidence that knowing others will protest weakens one's own temptation to do so \citep{cantoniEtAl2019}.

Taken together, this literature establishes both coordination and free-riding effects but leaves an important empirical question unresolved: how does the similarity of agents' information shift the balance between them? Field evidence cannot isolate this mechanism because changes in similarity are typically bundled with changes in the content, salience, reach, and network structure of information. Our experiment addresses this identification problem. We hold each agent's signal accuracy fixed while varying how similar agents' information is, and independently manipulate the threshold for collective success. This design identifies when more similar information shifts participation toward coordination and when it shifts participation toward free-riding.

Our experiment also connects three experimental literatures. A long line of work on threshold public goods has shown that coordination failures are common, and contributions depend on the location of the provision point and the return to reaching it \citep{palfreyRosenthal1984,cadsbyMaynes1999}. A second literature on turnout and participation games has established that participation tracks the perceived probability of being pivotal. \citet{levinePalfrey2007} showed that a quantal-response account organizes turnout in the laboratory, \citet{schramSonnemans1996} documented how participation responds strategically to group size and stakes, and \citet{duffyTavits2008} elicited pivotality beliefs directly with a proper scoring rule and found that these beliefs predict the decision to vote. Whereas the existing literature primarily compares subjective and objective pivotality in levels, we examine whether an exogenous change in the information structure moves perceived and actual pivotality together. A third literature studies how the structure of information shapes coordination. Experiments on global games find that public and private signals of the same precision have different effects because public information is more informative about what others will do \citep{heinemannNagelOckenfels2004}. Subsequent work measures this overweighting of public information and traces it to subjects' higher-order beliefs \citep{cornandHeinemann2014}, while experiments on information design study how control over the signal structure moves behavior \citep{frechetteLizzeriPerego2022}. The theory of information similarity, however, calls for bringing these strands together, which our experiment does.

Finally, our paper contributes to the literature on information in games. Economists have long studied how information affects private incentives and social outcomes \citep{hirshleifer1971}. In strategic environments, a large literature analyzes how information shapes equilibrium behavior and welfare \citep{morrisShin2002,angeletosPavan2007,bergemannMorris2013}, and a related comparative-statics literature obtains results in settings with monotone best responses or monotone equilibrium strategies \citep{vanZandtVives2007,jensen2018,mekonnenVizcaino2022}. However, a collective action game combines coordination and free-riding motives, so best responses are nonmonotonic. Moreover, much of this literature studies the arrival of new public information rather than changes in how \emph{similar} agents' information is. 

To compare environments by similarity, one needs an order over multivariate information structures. The literature offers several measures of interdependence \citep{mullerStoyan2002,meyerStrulovici2012}, alongside work on how moving from independent to dependent signals changes the value of information \citep{clemenWinkler1985,chengBorgers2024}, the set of feasible joint beliefs \citep{deOliveiraEtAl2023,arieliEtAl2021}, and the prospects for common learning \citep{crippsEtAl2008,awayaKrishna2022}. However, standard orders of interdependence do not generally imply that, after observing a given signal, an agent considers it more likely that others observed the same signal. Building on \citet{meyer1990}, \citet{basakDebKuvalekarGames} take a step beyond this framework: they introduce \emph{concentration along diagonal} (CAD), an order that ranks information structures by how similar agents' information is while holding fixed each agent's marginal informativeness. \citet{basakDebKuvalekarCollective} show why this new comparison matters for collective action. More similar information can either increase or decrease participation, depending on whether expected participation by others makes one's own action pivotal or redundant. Our experiment provides the first direct test of this comparative static.

The remainder of the paper proceeds as follows. Section~\ref{sec:theory} lays out the theoretical background and the behavioral hypotheses. Section~\ref{sec:experiment} describes the experiment. Section~\ref{sec:results} presents the results on reporting, beliefs, removal, and welfare, together with robustness checks. Section~\ref{sec:conclusion} concludes.

\section{Theoretical Background}\label{sec:theory}

This section sets out the theory the experiment tests. We describe the collective-action environment of \citet{basakDebKuvalekarCollective}, make precise what it means for information to become more similar, and show how the experimental parameter $\rho$ implements that notion while holding each agent's own information fixed. The central comparative static links the effect of more similar information to the difficulty of the collective goal: more similar information raises participation when the goal is demanding and lowers it when it is easy, depending on whether others' participation makes one's own action pivotal or redundant.

\subsection{Collective Action and Information Similarity}

We test the theory of information similarity in collective action developed by \citet{basakDebKuvalekarCollective}. An uncertain state $\theta\in\Theta=\{0,1\}$ determines whether collective action is desirable: when $\theta=1$ a change to the status quo benefits everyone, and when $\theta=0$ it does not. Agents simultaneously choose whether to take a costly action, and the action succeeds---the status quo is overturned---only if aggregate participation reaches a threshold. \citet{basakDebKuvalekarCollective} cast this as a regime-change game in which the threshold itself is uncertain, capturing the resilience of the status quo, and develop applications to protests, costly voting in committees, and public-good provision. Success when change is desirable delivers a public benefit to every agent, whether or not she participated, while participants pay a private cost in all states. Agents do not observe $\theta$; each receives a signal informative about it.

Because the benefit is public and the cost private, an agent participates only when she believes both that change is likely beneficial and that her own action is likely to be \emph{pivotal}---to determine whether the threshold is met. Beliefs about others therefore cut both ways: expecting greater participation by others can make one's own action more likely to be pivotal, strengthening the coordination motive, or more likely to be redundant, strengthening the free-riding motive. Which force dominates is a property of the environment, not of the information structure.

Holding that environment fixed, \citet{basakDebKuvalekarCollective} ask how the \emph{similarity} of agents' information shifts the balance between these forces. Since participation hinges on beliefs about others, what matters is not only how informative a signal is about $\theta$ but what it reveals about \emph{others'} signals. An agent faces two kinds of uncertainty: \emph{fundamental uncertainty} about the state, and \emph{strategic uncertainty} about what others have learned. To vary the second while holding the first fixed, \citet{basakDebKuvalekarCollective} order information structures by their \emph{concentration along diagonal} (CAD): one structure is more similar than another if, conditional on her own signal, an agent regards it as more likely that others observed the same signal, while every agent's marginal signal distribution is held fixed. More similar information thus leaves what a signal reveals about the world untouched and changes only what it implies about others. They state the order in two forms.

\begin{definition}[Concentration along diagonal; \citealp{basakDebKuvalekarCollective}]\label{def:cad}
Let $\mathcal{Y}\subset\mathbb{R}$ be finite, and let $Y$ and $\widehat{Y}$ be $\mathcal{Y}^2$-valued exchangeable signal pairs with distributions $\mathcal{D}$ and $\widehat{\mathcal{D}}$. Then $Y$ is \emph{more similar} than $\widehat{Y}$ in the CAD order, $\mathcal{D}\succeq_{\mathrm{CAD}}\widehat{\mathcal{D}}$, if
\begin{enumerate}
\item $Y_i$ and $\widehat{Y}_i$ are identically distributed for $i\in\{1,2\}$; and
\item for every $y\in\mathcal{Y}$ and $T\subseteq\mathcal{Y}$,
\begin{enumerate}[label=(\alph*)]
\item $\mathcal{D}(Y_2\in T\mid Y_1=y)\geq\widehat{\mathcal{D}}(\widehat{Y}_2\in T\mid \widehat{Y}_1=y)$ if $y\in T$, and
\item $\mathcal{D}(Y_2\in T\mid Y_1=y)\leq\widehat{\mathcal{D}}(\widehat{Y}_2\in T\mid \widehat{Y}_1=y)$ if $y\notin T$.
\end{enumerate}
\end{enumerate}
\end{definition}

\noindent Condition 1 fixes each agent's marginal; condition 2 says that observing $y$ makes any event containing $y$ more likely, and any event excluding it less likely, to describe another agent's signal. For a group of more than two agents, the order is restated through the distribution of the number of others who share one's signal.

\begin{definition}[Concentration along diagonal with $G>2$ agents; \citealp{basakDebKuvalekarCollective}]\label{def:cad-group}
Let $S=(S_1,\ldots,S_G)\in\mathcal{S}^{G}$ denote the signal vector, where $\mathcal{S}=\{0,1\}$. Let $\mathcal{P}\in\Delta(\Theta\times\mathcal{S}^{G})$ denote the joint distribution of $(\theta,S)$, and let $\mathcal{P}^{\theta}\in\Delta(\mathcal{S}^{G})$ denote the joint distribution of $S$ conditional on state $\theta$. Let $I_{-i}=\sum_{j\neq i}\mathbf{1}\{S_j=1\}$ denote the number of agents other than $i$ who receive signal $1$. Let $\gamma$ and $\widehat{\gamma}$ be the distributions of $I_{-i}$ conditional on $\theta=1$ and $S_i=1$. We say $\mathcal{P}^{1}\succeq_{\mathrm{CAD}}\widehat{\mathcal{P}}^{1}$ if there exists $k^{*}\in\{0,1,\ldots,G-2\}$ such that
  \[
    \gamma(k)\leq\widehat{\gamma}(k)
    \quad\text{for all } k\leq k^{*},
    \qquad
    \gamma(k)\geq\widehat{\gamma}(k)
    \quad\text{for all } k>k^{*}.
  \]
We call $k^{*}$ the index of sign change between $\mathcal{P}^{1}$ and $\widehat{\mathcal{P}}^{1}$.
\end{definition}

\noindent Thus, conditional on $\theta=1$ and $S_i=1$, more similar information shifts probability from realizations with at most $k^{*}$ other agents receiving signal $1$ toward realizations with more than $k^{*}$ other agents receiving signal $1$.

\citet{basakDebKuvalekarCollective} distinguish two environments in the following way. In an \emph{encouragement} environment~(E) an agent becomes more likely to be pivotal as participation rises, so that actions are \emph{strategic complements}; in a \emph{discouragement} environment~(D) she becomes more likely to be pivotal as participation falls, so that actions are \emph{strategic substitutes}. This dichotomy signs the effect of similarity. Let $\mathcal{V}^{*}(\mathcal{P})$ denote the maximal equilibrium aggregate participation under information structure $\mathcal{P}$---the largest expected number of participants, in the state where change is desirable, that any equilibrium sustains. The full equilibrium set cannot be characterized without further restrictions, but the effect of similarity on $\mathcal{V}^{*}$ is clear. Holding the marginal signal distributions fixed, their comparative static is \citep[Theorems 1--2]{basakDebKuvalekarCollective}:
\[
  \mathcal{P}^{\theta}\succeq_{\mathrm{CAD}}\widehat{\mathcal{P}}^{\theta}
  \ \text{for all }\theta
  \quad\Longrightarrow\quad
  \begin{cases}
    \mathcal{V}^{*}(\mathcal{P})\geq\mathcal{V}^{*}(\widehat{\mathcal{P}}), & \text{in (E)},\\[3pt]
    \mathcal{V}^{*}(\mathcal{P})\leq\mathcal{V}^{*}(\widehat{\mathcal{P}}), & \text{in (D)}.
  \end{cases}
\]
The second inequality holds under a regularity condition on the maximal equilibrium. Intuitively, more similar information promotes participation when participation by others makes an agent more likely to be pivotal, but discourages participation when it makes her own action more likely to be redundant.

\subsection{Information Similarity in a Content Moderation Game}

We study collective action in a content moderation game. In each round, a group of subjects faces a piece of content that is either harmful or benign. They do not observe the content itself; each receives an algorithmic report about it and decides whether to incur a cost to report it, and the content is removed only if reports reach a threshold $T$. Removing harmful content benefits every subject, reporter or not, so a report is a costly contribution to a public good. This is a collective action problem, exactly the kind studied in \citet{basakDebKuvalekarCollective}: success requires enough costly individual actions, the benefit is shared, and each subject's incentive depends on whether her own report is pivotal. 

Because the group is fixed and the threshold $T$ is commonly known, a subject who chooses to report harmful content is pivotal exactly when $T-1$ of the other eleven subjects also report. The likelihood of that pivotal event depends on a parameter $\rho$, defined below, that governs how similar subjects' algorithmic reports are. Writing $s_i$ for her signal, her incentive to report is the probability that the content is harmful, times the probability that her report is pivotal, times the benefit $B$ of removal, net of the cost $c$ of reporting. That is,
\[
    \Pr(\theta=1\mid s_i)\,\Pr(A_{-i}=T-1\mid s_i,T,\rho)\,B-c.
\]
The first term is fundamental information about the state; the second is perceived pivotality. Information similarity, which $\rho$ controls, operates entirely on this second term.

The information structure is where we implement similarity. Each algorithmic evaluation produces one of two reports: a warning (\textsc{Likely Harmful}), denoted by $W$, or \textsc{Likely Benign}. When the content is harmful, the evaluation produces $W$ with probability $0.75$ and \textsc{Likely Benign} with probability $0.25$. When the content is benign, it always produces \textsc{Likely Benign}. Thus, the algorithm may fail to flag harmful content but never falsely flags benign content; observing $W$ therefore reveals that the content is harmful, so $\Pr(\theta=1\mid W)=1$. We vary the correlation of reports across subjects. With probability $\rho$, the algorithm performs a single evaluation of the content and sends the resulting report to every subject. With probability $1-\rho$, it performs independent evaluations for each subject and privately sends each subject the result of her own evaluation. Raising $\rho$ leaves the accuracy of each subject's own report unchanged but makes her report more predictive of the reports received by others. Specifically, conditional on the content being harmful and subject $i$ receiving a warning, the probability that another subject $j$ also receives a warning is
\[
    \Pr(s_j=W\mid s_i=W,\rho)=0.75+0.25\rho.
\]
Thus, raising $\rho$ makes information more similar in the CAD order---more similar information with marginals held fixed (Definitions~\ref{def:cad}--\ref{def:cad-group}).

Because our game is finite, with binary signals and a commonly known threshold, the theory's comparative static specializes to a single switch point. Following \citet{basakDebKuvalekarCollective}, we restrict attention to symmetric pure strategies and let $\sigma^{1}$ denote the strategy of reporting if and only if warned; their result further assumes that the cost is high enough that an unwarned subject never reports. Let $\gamma_\rho(k)$ be the probability that exactly $k$ of the other eleven subjects report, conditional on harmful content and a warning. More similar information shifts $\gamma_\rho$ toward higher counts, and the low- and high-$\rho$ distributions cross exactly once, at the index $k^{*}$ of Definition~\ref{def:cad-group}. Let $\mathcal{E}(\mathcal{P})$ denote the set of pure-strategy equilibria under information structure $\mathcal{P}$.

\begin{proposition}[\citealp{basakDebKuvalekarCollective}, Proposition~4]\label{prop:switch}
Fix the threshold $T$. Suppose $\mathcal{P}^{1}\succeq_{\mathrm{CAD}}\widehat{\mathcal{P}}^{1}$, and let $k^{*}$ be the associated index of sign change between $\mathcal{P}^{1}$ and $\widehat{\mathcal{P}}^{1}$. Suppose $\sigma^{1}\in\mathcal{E}(\widehat{\mathcal{P}})$.
\begin{enumerate}
  \item If $k^{*}<T-1$, then $\sigma^{1}\in\mathcal{E}(\mathcal{P})$.
  \item If $k^{*}\ge T-1$, then it is possible that $\sigma^{1}\notin\mathcal{E}(\mathcal{P})$.
\end{enumerate}
\end{proposition}

\noindent The mechanism is the pivotality $\gamma_\rho(T-1)$: more similar information raises it when the pivotal count lies above the crossing, $T-1>k^{*}$, and lowers it below, so the threshold $T^{*}=k^{*}+1$ separates the discouragement and encouragement environments. Our experiment uses this result as a benchmark without imposing $\sigma^{1}$ on observed behavior. By varying $\rho$ and $T$, we test whether threshold-dependent changes in perceived pivotality organize reporting and whether the effect of similarity changes sign. Our individual-level test conditions on a warning, which reveals the state. The theory thus predicts a single sign reversal in the effect of similarity. Where that reversal occurs in our environment is an empirical question, because $k^{*}$ depends on the reporting rate sustained in equilibrium.

This logic maps onto the threshold $T$, which we vary across rounds. The pivotal event is always ``exactly $T-1$ of the other eleven subjects report,'' but what changes across thresholds is whether more similar information makes that event more or less plausible. When $T$ is high, a subject is pivotal only if many others report, so more similar information should raise reporting by making a coordinated wave of reports more plausible; this is an encouragement environment. When $T$ is low, she is pivotal only if few others report, so more similar information should lower it by making it likely that others' reports already suffice; this is a discouragement environment.

These predictions lead to two empirical questions. First, at which threshold in our environment does the effect of information similarity on reporting change sign? Second, how do the resulting changes in individual reporting translate into removal of harmful content and welfare? A change in reporting affects removal only if it moves the group across the threshold $T$; otherwise, it affects welfare only through reporting costs. The next subsection states the corresponding hypotheses, and Section~\ref{sec:experiment} describes how we vary $\rho$ and $T$ independently to test them.

\subsection{Behavioral Hypotheses}

The predictions above imply three hypotheses. The first two concern a warned subject ($s_i=W$). A warning reveals the state ($\Pr(\theta=1\mid W)=1$), so more similar information shifts only her beliefs about others, not about whether the content is harmful. The third concerns how individual reporting aggregates into collective outcomes. 

Let $R_i$ denote subject $i$'s reporting decision, and let $\pi_i(T)$ denote her elicited probability that exactly $T-1$ other subjects report. We define the full-range effects of information similarity at threshold $T$ as
\[
  \begin{aligned}
  \Delta_R(T)
  &=
  \mathbb{E}[R_i\mid \rho=1,T,s_i=W]
  -
  \mathbb{E}[R_i\mid \rho=0,T,s_i=W],\\
  \Delta_P(T)
  &=
  \mathbb{E}[\pi_i(T)\mid \rho=1,T,s_i=W]
  -
  \mathbb{E}[\pi_i(T)\mid \rho=0,T,s_i=W].
  \end{aligned}
\]
Thus, $\Delta_R(T)$ is the change in a warned subject's reporting probability, and $\Delta_P(T)$ is the change in her perceived probability of being pivotal, as information moves from fully idiosyncratic to fully common at threshold $T$. Empirically, we estimate each as the slope of the outcome on $\rho$ over the full grid $\rho\in \{0,0.1,\dots,1\}$.

\begin{hypothesis}[Reporting reversal]\label{hyp:crossover}
The effect of information similarity on reporting changes sign once as the threshold rises: there is a cutoff $T^{*}$ such that
\[
    \Delta_R(T)<0 \ \text{ for } T<T^{*},
    \qquad
    \Delta_R(T)>0 \ \text{ for } T>T^{*}.
\]
The theory predicts a single cutoff; we will locate it in our environment empirically.
\end{hypothesis}

\begin{hypothesis}[Pivotality channel]\label{hyp:pivotality}
The effect of information similarity on perceived pivotality changes sign at the same cutoff as its effect on reporting: 
\[
    \Delta_P(T)<0 \quad \text{for } T<T^{*},
    \qquad
    \Delta_P(T)>0 \quad \text{for } T>T^{*}.
\]
More similar information raises the expected number of other reporters in every regime; what reverses is whether this leaves the subject more likely to be pivotal or more likely to be redundant. A warned subject is more likely to report when she perceives a greater chance that her report will be pivotal. Thus, perceived pivotality and reporting move in the same direction on either side of the cutoff.
\end{hypothesis}

\begin{hypothesis}[Collective outcomes]\label{hyp:collective}
When the threshold is reachable, similarity-induced changes in reporting move the probability of removing harmful content in the same direction; when the threshold is out of reach, additional reporting leaves removal unchanged. Welfare, defined as the public benefit from removal net of reporting costs, tracks removal when the threshold is reachable. When the threshold is out of reach, extra reporting adds costs without increasing removal, so more similar information can raise reporting while lowering welfare.
\end{hypothesis}

\section{The Experiment}\label{sec:experiment}

We design a laboratory environment that implements the central elements of the collective-action problem. An uncertain state determines whether collective action is desirable. Participation is individually costly, while collective success generates a public benefit when enough subjects participate. Subjects receive informative signals whose similarity can vary independently of their marginal informativeness. The theory provides a benchmark for how information similarity should affect participation: more similar information should discourage participation when only a few subjects are needed and encourage it when many are needed. The experiment asks whether this reversal appears in behavior, whether it operates through beliefs about being pivotal, and whether the resulting changes in participation improve collective success and welfare.

\subsection{Environment}

We implement the collective-action problem as a content-moderation game. In each round, subjects are randomly rematched into groups of twelve, and each group faces a piece of content that is harmful with probability $0.8$ and benign with probability $0.2$. Each subject receives an algorithmic report about the content and then chooses to \textbf{Report} it, at a cost of $30$ points, or to \textbf{Ignore} it, at no cost. Writing $A$ for the number of reports in her group, the content is removed exactly when $A\geq T$. Removing harmful content pays every member $100$ points, whether or not she reported; removing benign content, or leaving harmful content up, pays nothing. Table \ref{tab:experimental-payoff} collects these payoffs.

\begin{table}[!htbp]
\centering
\caption{Payoff Structure in the Experiment}
\label{tab:experimental-payoff}
\begin{tabular}{lccc}
\toprule
& \multicolumn{2}{c}{Content removed ($A\geq T$)}
& Content not removed ($A<T$) \\
\cmidrule(lr){2-3}
& Harmful content & Benign content & Any content \\
\midrule
\textbf{Report} & $100-30=70$ & $0-30=-30$ & $-30$ \\
\textbf{Ignore} & $100$ & $0$ & $0$ \\
\bottomrule
\end{tabular}
\caption*{\footnotesize \textit{Notes}: Payoffs are in points. One randomly selected paid round is converted to cash at 100 points = CNY 50.}
\end{table}

The table makes reporting strategically nontrivial. Reporting pays off only when the content is harmful and the subject's own report is what carries the group over the threshold: if enough others report, ignoring earns the same benefit for free; if too few do, reporting simply burns $30$ points. The decision therefore depends on the chance that exactly $T-1$ others report---the event in which the subject is pivotal.

\subsection{Treatments}

\subsubsection{Treatment 1: Information Similarity}

Before deciding, each subject receives one of two algorithmic reports: \textsc{Likely Harmful} or \textsc{Likely Benign}. If the content is benign, the algorithm always produces \textsc{Likely Benign}. If the content is harmful, it produces \textsc{Likely Harmful} with probability $0.75$ and \textsc{Likely Benign} with probability $0.25$.

In each group-round, a similarity parameter $\rho$ is drawn uniformly at random from the eleven values $\{0,0.1,\ldots,1\}$ and displayed to all subjects. With probability $\rho$, the algorithm evaluates the content once and sends the resulting report to every group member. With probability $1-\rho$, it performs twelve independent evaluations of the content, one for each subject, and privately sends each subject the result of her own evaluation. Every evaluation follows the same signal technology described above. Raising $\rho$ therefore leaves the accuracy of each subject's own report unchanged but makes it more informative about the reports received by others. This is a CAD increase that holds marginal information fixed (Definitions~\ref{def:cad}--\ref{def:cad-group}).

\subsubsection{Treatment 2: Removal Threshold}

Each session is assigned to one of three threshold regimes, and within a session the exact threshold $T$ is drawn afresh each group-round, uniformly at random from that regime's four thresholds (Table~\ref{tab:design-regimes}). Both $\rho$ and $T$ are drawn independently across the four groups in each session-round. Across the three regimes, the threshold spans its full feasible range, from one report to unanimity. Subjects observe the realized $\rho$ and $T$ before making their decisions.

\begin{table}[!htbp]
\centering
\caption{Threshold regimes}
\label{tab:design-regimes}
\begin{tabular}{lcc}
\toprule
Regime & Threshold values & Sample size \\
\midrule
Low    & $T\in\{1,2,3,4\}$    & 4 sessions, 192 subjects \\
Medium & $T\in\{5,6,7,8\}$    & 4 sessions, 192 subjects \\
High   & $T\in\{9,10,11,12\}$ & 4 sessions, 192 subjects \\
\bottomrule
\end{tabular}
\end{table}

We vary information similarity $\rho$ within subjects across its full range but assign the threshold regime between sessions. The asymmetry reflects the two parameters' different roles and mirrors the field. Similarity is the treatment and the naturally fluctuating one: under a fixed institution, how correlated people's information is changes from one event to the next, so within-subject variation in $\rho$ reproduces that counterfactual and identifies its effect free of individual heterogeneity. The threshold, by contrast, is the institution that converts actions into collective success---an electoral or quorum rule, a platform's content-moderation policy, an organizational decision rule---and such institutions are persistent. Holding the regime fixed lets each subject reason within a stable environment: a threshold swinging from one report to near-unanimity round to round would make the strategic problem non-stationary and invite the carry-over and demand effects that afflict within-subject designs spanning many parameter values \citep{charnessGneezyKuhn2012}. The four exact thresholds drawn within each regime nonetheless provide within-subject variation that identifies the threshold's local effect.

\subsection{Belief Elicitation}

Before choosing whether to \textbf{Report} or \textbf{Ignore}, each subject answers two incentivized belief questions. The first elicits her belief about aggregate participation; the second elicits her perceived probability of being pivotal.

\paragraph{Predicted participation.}
The first question asks the subject to predict how many of the other eleven group members will report. She receives CNY~10 if her prediction is exactly correct, CNY~5 if it differs from the realized number by one, and nothing otherwise.

\paragraph{Perceived pivotality.}
The second question asks the subject to report, on a scale from 0 to 100 percent, the probability that exactly $T-1$ other group members will report. This is precisely the event in which her own report is pivotal. We incentivize the answer using a Becker--DeGroot--Marschak mechanism \citep{bdm1964}. The reported probability $X$ is compared with a uniformly drawn number $Y\in\{0,\ldots,100\}$. If $X>Y$, the subject receives CNY~10 if the pivotal event occurs and nothing otherwise; if $X\leq Y$, she instead receives a lottery that pays CNY~10 with probability $Y/100$. Truthfully reporting her perceived probability maximizes her chance of receiving the bonus.

The two questions measure distinct beliefs. A subject may expect many others to report without believing that exactly $T-1$ will do so. At a high threshold, expecting more reporters can move participation toward the pivotal count; at a low threshold, it can move participation beyond that count and make her own report redundant. Separating the two beliefs therefore allows us to distinguish beliefs about aggregate participation from beliefs about whether one's own action will matter.

\begin{figure}[!htbp]
\centering
\includegraphics[width=0.55\textwidth]{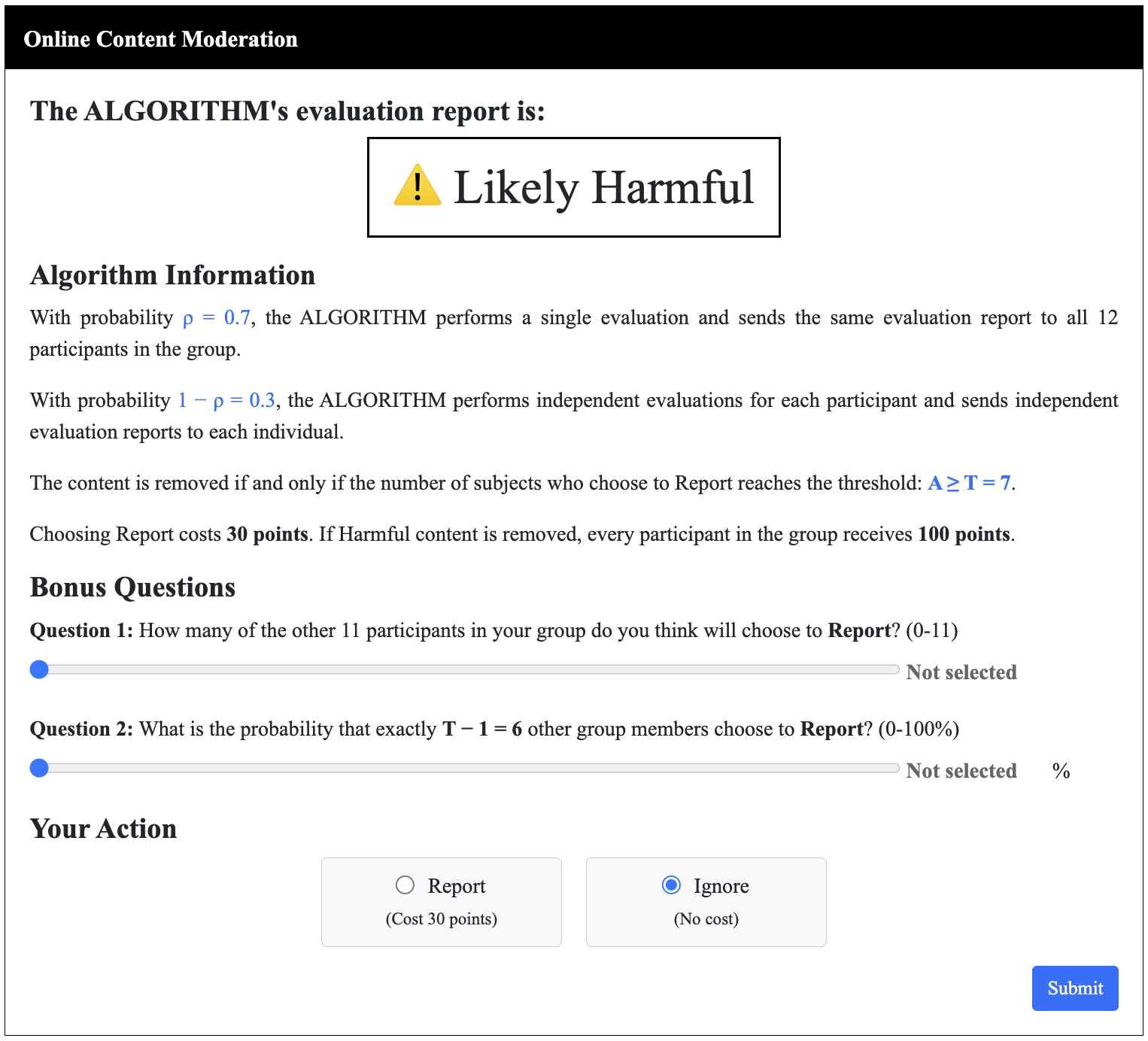}
\caption{The decision screen}
\label{fig:decision-screen}
\caption*{\footnotesize \textit{Notes}: A subject's screen in a paying round, showing her algorithmic report (\textsc{Likely Harmful} or \textsc{Likely Benign}), the realized similarity parameter $\rho$ and threshold $T$, the two belief questions, and the \textsc{Report}/\textsc{Ignore} choice. The experiment was conducted in Chinese; the figure shows an English rendering of the screen.}
\end{figure}

Figure~\ref{fig:decision-screen} shows the screen on which subjects observe their algorithmic report, $\rho$, and $T$, answer both belief questions, and make their reporting decision.

\subsection{Procedures and Sample}

The experiment was programmed in oTree \citep{chenSchongerWickens2016} and conducted in Chinese at Wuhan University's Research Center for Behavioral Science in May and June 2026. We ran twelve sessions with $48$ subjects each, for a total of $576$ subjects. Each session was assigned to one threshold regime, giving four sessions and $192$ subjects per regime. Sessions lasted approximately 90 minutes.

At the beginning of each session, subjects received printed instructions, which an experimenter read aloud. They then completed a comprehension quiz covering the signal-generation process, threshold rule, payoffs, and belief-elicitation
procedures, and were required to answer every question correctly before proceeding.\footnote{To encourage careful reading, subjects were informed in advance that answering every question correctly on their first attempt would earn an additional CNY~5.} Subjects next completed three unpaid practice rounds followed by twenty main rounds, one of which was randomly selected for payment. In each main round, the $48$ subjects were randomly rematched into four groups of twelve. After every main round, subjects learned the true state, the number of reports in their group, whether the content was removed, and their payoff for that round. After the main task, subjects completed an end survey containing incentivized measures of risk attitudes, altruism, and trust, followed by non-incentivized questions about their decision-making and experience in the experiment.

Payment consisted of a CNY~20 show-up fee; the decision payoff and both belief-elicitation bonuses from one randomly selected paid round; the CNY~5 comprehension-quiz bonus, when earned; and earnings from the incentivized end-survey tasks. Experimental points were converted at a rate of 100 points to CNY~50. Average earnings were CNY~54.30. The full instructions, comprehension quiz, and end survey appear in Appendix~\ref{app:materials}, Sections~\ref{app:instructions}--\ref{app:end_survey}. Figure~\ref{fig:session-flow} summarizes the sequence of events within an experimental session.

\begin{figure}[!htbp]
\centering
\begin{tikzpicture}[
stage/.style={
    draw,
    align=center,
    text width=2.6cm,
    minimum height=2.6cm,
    inner sep=4pt,
    font=\small
},
flow/.style={
    -{Stealth[length=5pt,width=5pt]},
    line width=0.8pt
}
]

\node[stage] (instructions) at (0,0)
{Printed\\Instructions\\Read Aloud};

\node[stage] (quiz) at (3.2,0)
{Comprehension\\Quiz};

\node[stage] (practice) at (6.4,0)
{3 Unpaid\\Practice Rounds};

\node[stage] (main) at (9.6,0)
{20 Main Rounds\\with Random Rematching\\and Round Feedback};

\node[stage] (survey) at (12.8,0)
{End Survey\\and Payment};

\draw[flow] (instructions) -- (quiz);
\draw[flow] (quiz) -- (practice);
\draw[flow] (practice) -- (main);
\draw[flow] (main) -- (survey);

\end{tikzpicture}

\caption{Flow of an experimental session}
\label{fig:session-flow}
\caption*{\footnotesize \textit{Notes}: The figure summarizes the sequence of a session. One of the twenty main rounds was randomly selected to determine the subject's decision payoff and two belief-elicitation bonuses.}
\end{figure}

The experiment generated $11,520$ individual decisions and $960$ group-round observations. Tests of the individual-level mechanism use the \emph{warning sample}: the $7,035$ decisions in which a subject received \textsc{Likely Harmful}.\footnote{In the complementary \emph{likely-benign} sample---the remaining $4{,}485$ decisions---reporting is uniformly low, $6.35\%$ overall and between $4.99\%$ and $7.72\%$ across the three regimes. These decisions carry little strategic content; Appendix~\ref{app:additional-results} reports them in full.} Because this warning occurs only when the content is harmful, the warning sample fixes the state and allows $\rho$ to shift beliefs about others without changing beliefs about whether the content is harmful. The individual outcomes are reporting and the two elicited beliefs; the group outcomes are the number of reports, removal of harmful content, and realized welfare.

\section{Results}\label{sec:results}

We study how information similarity affects costly reporting as the removal threshold rises, how the elicited beliefs covary with the reporting response, whether those beliefs are well calibrated, and how the resulting reporting decisions translate into removal and welfare. Recall that, in each twelve-person group, $\rho$ is the probability that subjects' algorithmic reports come from a common evaluation rather than separate independent evaluations, while $T$ is the number of reports required for removal. The individual-level analyses use decisions made after a warning, which reveals that the content is harmful; within this sample, variation in $\rho$ changes what subjects infer about others without changing what they know about the state.

The results reveal a tension between individual participation and collective outcomes. At the individual level, the effect of similarity on reporting reverses once as the removal threshold rises (Result~\ref{res:crossover}), and perceived pivotality moves with this reversal (Result~\ref{res:pivotality}). Subjects overestimate pivotality in all three regimes, with experience narrowing the gap only in the High regime (Result~\ref{res:calibration}). Beyond this level miscalibration, however, perceived pivotality responds to information similarity in line with actual pivotality only in the Medium regime: the Low regime exhibits unrecognized pivotality, the Medium regime aligned pivotality, and the High regime illusory pivotality (Result~\ref{res:alignment}). At the group level, changes in reporting translate into changes in harmful-content removal only when the threshold is within reach (Result~\ref{res:removal}). Consequently, the effect of similarity on welfare is negative in the Low regime, positive in the Medium regime, and negative again in the High regime (Result~\ref{res:welfare}). Its welfare consequences therefore depend not only on whether similarity mobilizes reporting, but also on whether the institutional threshold converts those reports into removal.

\subsection{Locating the Reporting Crossover}\label{subsec:reporting-crossover}

\begin{result}[The reporting crossover]\label{res:crossover}
The effect of more similar information on reporting changes systematically with the threshold regime. It lowers reporting in the Low regime, raises reporting in the Medium regime, and raises reporting even more strongly in the High regime. The sign switches at an estimated threshold cutoff of $T^{*}\approx4.5$ reports in a twelve-person group in our environment.
\end{result}

\begin{figure}[!htbp]
\centering
\includegraphics[width=0.65\textwidth]{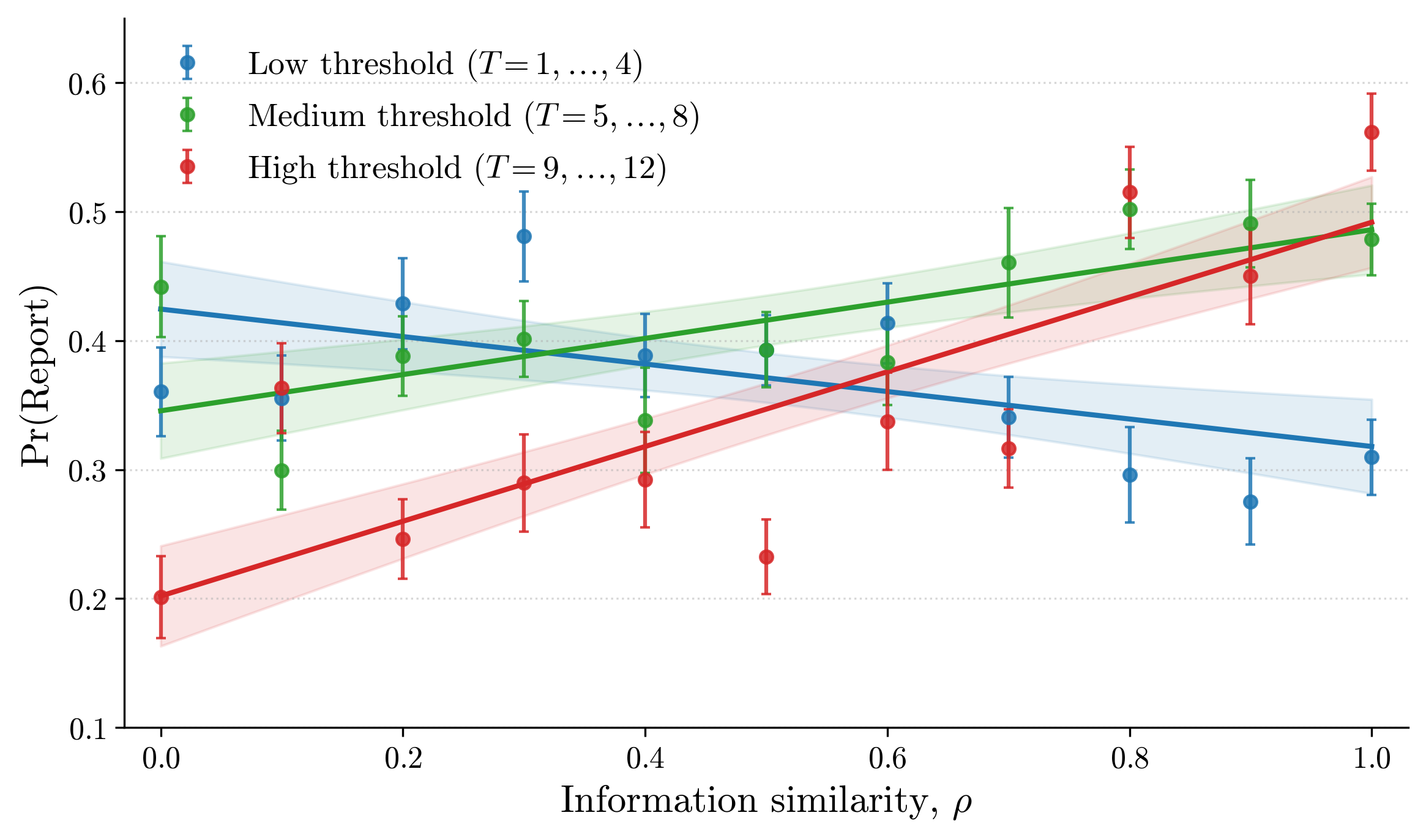}
\caption{Reporting against information similarity by threshold regime}
\label{fig:report-rho}
\caption*{\footnotesize \textit{Notes}: Markers show mean reporting at each realized value of $\rho$; lines are OLS fits with 95 percent confidence bands. The plotted sample consists of decisions made after the participant received a warning.}
\end{figure}

Figure~\ref{fig:report-rho} displays the reporting response in all three threshold regimes. As information becomes more similar, reporting declines in the Low regime, increases in the Medium regime, and increases more steeply in the High regime. Column~(1) of Table~\ref{tab:theoryreport} estimates the corresponding slopes as $-0.168$, $+0.156$, and $+0.341$. Thus, moving from fully idiosyncratic to fully common information reduces reporting by about 17 percentage points in the Low regime, increases it by about 16 points in the Medium regime, and increases it by about 34 points in the High regime. The High-minus-Low difference is $0.509$ (s.e.\ $0.054$; $p<0.001$).\footnote{This difference and the exact-threshold interaction below are robust to inference at the session level, the level at which the threshold regime is assigned: a wild cluster bootstrap over the twelve sessions and randomization inference over the assignment of sessions to regimes both leave them significant (Appendix~\ref{app:additional-results}, Table~\ref{tab:session-inference}).} This ordering supports the theory's central comparative static: more similar information discourages costly reporting when the collective threshold is easy to reach, encourages it when the threshold is more demanding, and has an increasingly positive effect as the threshold rises. The regime-level comparison therefore establishes the predicted reversal, but each regime pools four exact thresholds and cannot reveal where the similarity effect crosses zero. To identify the empirical cutoff $T^{*}$, we pool all sessions and estimate the similarity slope as a linear function of the exact threshold, whose full range is spanned across the between-session regime assignment.

\begin{table}[!htbp]
\centering
\caption{Information similarity, reporting, and pivotality beliefs.}
\label{tab:theoryreport}
\small
\begin{tabular}{@{}lccc@{}}
\toprule
 & (1) Report & (2) Pivotality & (3) Expected \\
 & & probability (pp) & other reporters \\
\midrule
Similarity $\rho$ (Low slope) & -0.168*** & -6.75*** & 0.626*** \\
 & (0.035) & (2.02) & (0.195) \\
$\rho\times$Medium & 0.324*** & 10.97*** & 0.954*** \\
 & (0.044) & (2.75) & (0.260) \\
$\rho\times$High & 0.509*** & 25.94*** & 2.586*** \\
 & (0.054) & (3.48) & (0.391) \\
Constant & 0.325*** & 24.40*** & 3.834*** \\
 & (0.011) & (0.71) & (0.081) \\
\addlinespace
\multicolumn{4}{@{}l}{\textit{Implied similarity slopes}} \\
\quad Medium & 0.156*** & 4.21** & 1.580*** \\
\quad High & 0.341*** & 19.19*** & 3.212*** \\
\midrule
Threshold, round, participant FE & Yes & Yes & Yes \\
Observations & 7,035 & 7,035 & 7,035 \\
$R^2$ & 0.437 & 0.433 & 0.469 \\
\bottomrule
\end{tabular}
\par\vspace{2pt}
\begin{minipage}{0.92\textwidth}
\footnotesize \textit{Notes:} Column (1) is a linear probability model for the reporting indicator; column (2) is the elicited probability (in percentage points) that exactly $T-1$ other group members report; column (3) is the elicited expected number of other reporters. All columns include threshold, round, and participant fixed effects. Low is the omitted regime, so $\rho$ is the Low-threshold similarity slope, and the implied Medium and High slopes add the corresponding interaction. Standard errors are two-way clustered by participant and group-round in parentheses; $^{*}p<0.1$, $^{**}p<0.05$, $^{***}p<0.01$. The estimation sample consists of decisions made after the participant received a warning.
\end{minipage}
\end{table}

The continuous interaction specification in Appendix~\ref{app:additional-results}, Table~\ref{tab:thresholdinteraction}, estimates the reporting slope at threshold $T$ as $-0.253+0.056\,T$. The interaction coefficient of $0.056$ (s.e.\ $0.006$; $p<0.001$) means that each one-unit increase in $T$ raises the similarity slope by 5.6 percentage points, so the slope is negative at low thresholds and turns positive as the threshold becomes more demanding---the single crossing the theory predicts, with a location that the model does not pin down (Section~\ref{sec:theory}). Setting the estimated slope equal to zero, $-0.253+0.056\,T^{*}=0$, gives $T^{*}\approx4.5$, or a crossing index of $k^{*}=T^{*}-1\approx3.5$ other reporters, a little over a third of the group. We do not read this value as a structural constant: the theory predicts that the reversal occurs once, and its location depends on the equilibrium reporting rate and on the specific thresholds, group size, and payoffs. What our design identifies is the existence and direction of the reversal, not a universal cutoff.

Having established where the reporting response changes sign, we next ask how the elicited beliefs relate to that reversal.

\subsection{Perceived Pivotality and the Reporting Reversal}

\begin{result}[Pivotality beliefs and reporting]\label{res:pivotality}
Perceived pivotality mirrors the reporting crossover, whereas expected participation rises in every regime. More similar information raises perceived pivotality where the threshold is demanding and lowers it where the threshold is easy, and stated pivotality is positively associated with reporting.
\end{result}

The two elicited beliefs differ in the pattern predicted by the theory. The first question asks how many of the other eleven members will report, and on this measure $\rho$ does what raising signal correlation should: after a warning, subjects expect more others to report when information is more similar, with a positive slope in every regime---about $0.63$ more expected reporters in the Low regime as $\rho$ moves from $0$ to $1$, and about $3.21$ in the High regime (Table~\ref{tab:theoryreport}, column~(3)). Part of that gap is mechanical, since more correlated signals mean a warned subject should expect more others to be warned; but the positive expected-participation response is common to all regimes and does not match the regime-dependent reporting reversal. One interpretation is that subjects also anticipate how others turn signals into actions: where many reports are needed, more similar information suggests a coordinated wave, while where few are needed it suggests others will let someone else act.

Expected participation, however, is not pivotality. The second question asks for the probability that \emph{exactly} $T-1$ others report---the event in which the subject's own report is pivotal---and here the sign flips with the threshold (Table~\ref{tab:theoryreport}, column~(2)). More similar information lowers perceived pivotality by $6.75$ percentage points in the Low regime, raises it by $4.21$ points in the Medium regime, and raises it by $19.19$ points in the High regime. The belief response mirrors the reporting response, regime for regime.

\begin{figure}[!htbp]
\centering
\includegraphics[width=\textwidth]{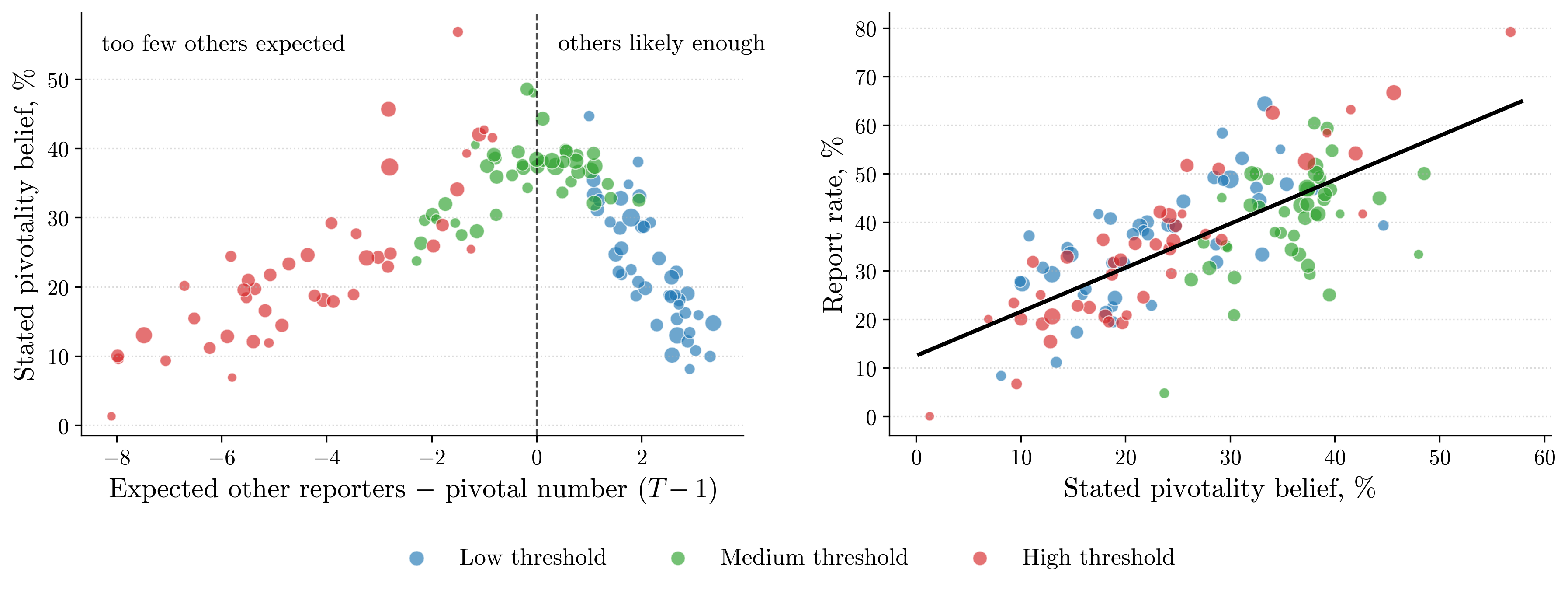}
\caption{Expected gap, pivotality beliefs, and reporting}
\label{fig:gap-pivotality-report}
\caption*{\footnotesize \textit{Notes}: The left panel plots the mean stated probability that exactly $T-1$ other subjects report against the mean expected number of other reporters minus $T-1$. The right panel plots cell report rates against stated pivotality beliefs; the solid black line is a linear fit. Marker size is proportional to the number of observations in the cell. The plotted sample consists of decisions made after the participant received a warning.}
\end{figure}

Figure~\ref{fig:gap-pivotality-report} shows the corresponding descriptive patterns in levels. The left panel plots stated pivotality against the gap between the expected number of other reporters and the pivotal count $T-1$: pivotality peaks when expected participation sits near the pivotal count and falls when it lands well short of or well past it. The right panel shows a positive association between the belief and behavior---cells with higher stated pivotality report more. This pattern is inconsistent with a purely imitative story. In the Low regime, subjects grow more confident that others will report while both perceived pivotality and reporting fall. More generally, the evidence is consistent with the theory's distinction between expected actions moving a subject toward versus away from the pivotal count.

Finally, stated pivotality is positively associated with reporting. In the controlled specification, a one-percentage-point higher stated pivotality is associated with a 0.48-percentage-point higher reporting probability (s.e.\ $0.03$; $p<0.001$). Adding stated pivotality reduces the High-minus-Low similarity slope from $0.509$ to $0.385$, an attenuation of 24.3 percent. Together, these patterns provide suggestive evidence that perceived pivotality partially mediates the reporting reversal.

\subsection{Pivotality Calibration and Alignment}

Result~\ref{res:pivotality} establishes that perceived pivotality moves with information similarity and predicts reporting. A distinct behavioral question is whether these beliefs are well calibrated: do subjects accurately assess how often their own report can change removal?

\begin{result}[Pivotality is systematically overestimated]
\label{res:calibration}
Subjects systematically overestimate how often their own report is pivotal in all three threshold regimes. The gap between perceived and actual pivotality is smallest in the Low regime and largest in the High regime. Experience improves calibration only in the High regime, and even there perceived pivotality remains substantially above its realized frequency.
\end{result}

\begin{figure}[!htbp]
\centering
\includegraphics[width=\textwidth]{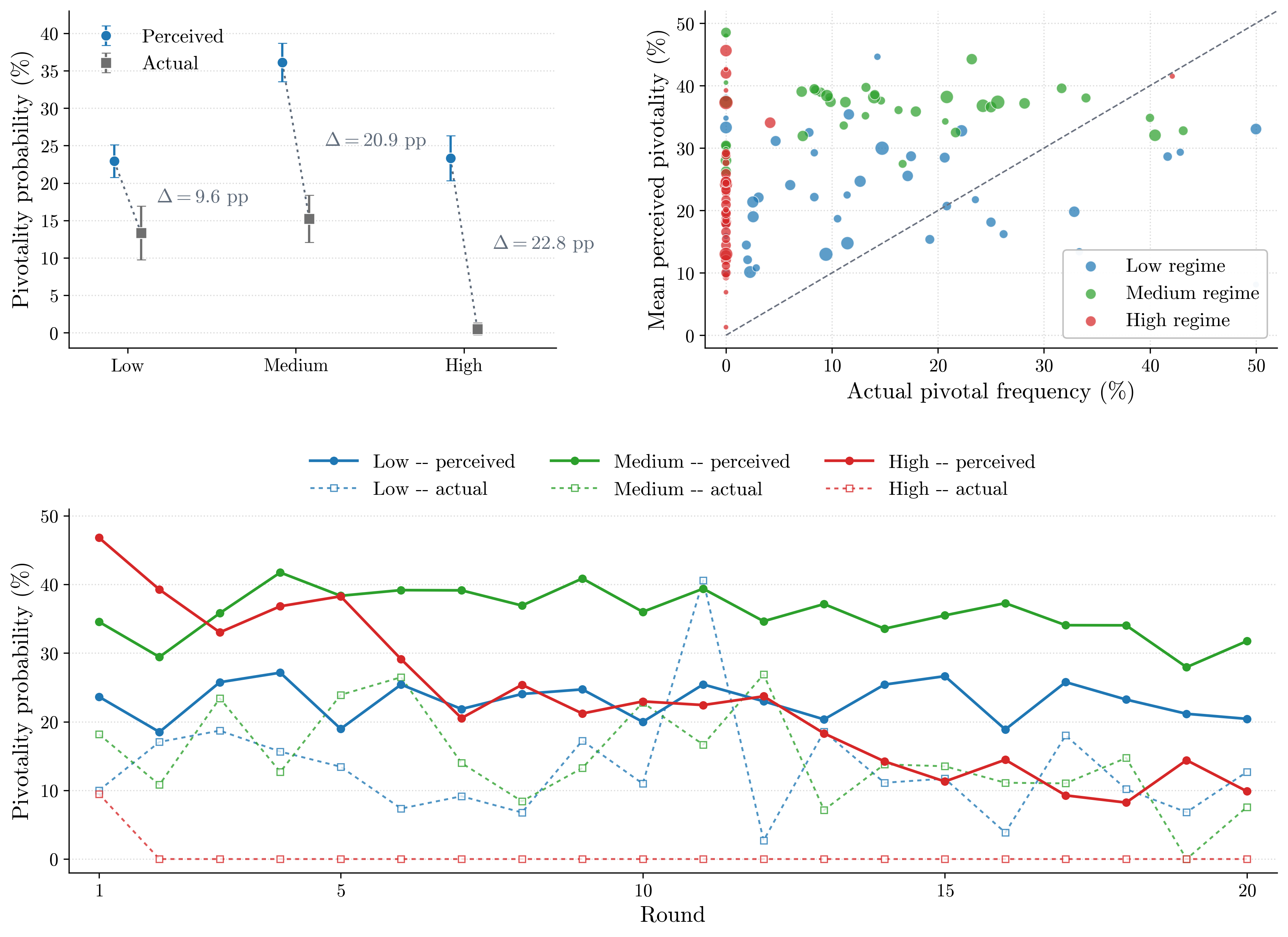}
\caption{Perceived and actual pivotality: calibration and learning}
\label{fig:pivotality-calibration}
\caption*{\footnotesize \textit{Notes}: Let $A$ denote total reports in the group and $R_i$ subject $i$'s reporting decision. Actual pivotality is the realized indicator $\mathbf{1}\{A-R_i=T-1\}$, which equals one exactly when changing subject $i$'s action would change removal. The top-left panel reports observation-level means with 95 percent confidence intervals; standard errors are two-way clustered by participant and group-round. In the top-right panel, observations are grouped into exact $(T,\rho)$ cells. Each marker compares the cell's mean stated probability with the realized pivotal frequency; marker size is proportional to the number of observations, and the dashed line is perfect calibration. The bottom panel plots observation-level means of perceived and actual pivotality by round and threshold regime. The early-to-late comparisons reported in the text control for participant and exact $(T,\rho)$-cell fixed effects; standard errors are two-way clustered by participant and group-round. The plotted sample consists of decisions made after the participant received a warning.}
\end{figure}

Across the warning sample, subjects overestimate pivotality by $17.6$ percentage points (s.e.\ $1.1$; $p<0.001$). The gap is smallest in the Low regime, where perceived pivotality is $22.9$ percent and actual pivotality is $13.3$ percent. It is more than twice as large in each of the two more demanding regimes. In the Medium regime, the corresponding values are $36.1$ and $15.2$ percent; in the High regime, subjects assign a $23.3$ percent probability to an event that occurs in only $0.5$ percent of decisions (see Appendix~\ref{app:additional-results}, Table~\ref{tab:pivotality-calibration}). Figure~\ref{fig:pivotality-calibration} also shows this pattern. The top-left panel displays the regime-level perceived and actual means, while the top-right panel compares the mean perceived probability with the realized pivotal frequency separately in each exact $(T,\rho)$ cell. Most cells lie above the 45-degree line, and the High-regime cells are concentrated near zero actual pivotality even when perceived pivotality remains substantial.

The preceding analysis already shows that perceived pivotality predicts reporting. To distinguish perceived from actual pivotality, we construct for each decision an empirical benchmark from other group-rounds with the same exact $(T,\rho)$, excluding the subject's current group-round, and enter it alongside the stated belief in the participant fixed-effects specification reported in Appendix~\ref{app:additional-results}, Table~\ref{tab:calibration-reporting}. A 10-percentage-point increase in perceived pivotality predicts a $4.7$-point increase in reporting (s.e.\ $0.3$; $p<0.001$), while a 10-point increase in the empirical actual-pivotality benchmark predicts only $0.2$ points (s.e.\ $0.5$; $p=0.669$). Reporting, therefore, follows perceived rather than empirical actual pivotality.

The calibration gap also changes with experience, but only where pivotality is nearly absent. The bottom panel of Figure~\ref{fig:pivotality-calibration} traces perceived and realized pivotality by round in all three regimes. In the Low and Medium regimes, the two series fluctuate without a sustained narrowing of the gap. In the High regime, by contrast, perceived pivotality trends downward: its mean falls from $31.8$ percent in rounds 1--10 to $14.7$ percent in rounds 11--20. Realized pivotality is $1.0$ percent in the first half and zero in the second, and the gap narrows from $30.8$ to $14.7$ percentage points rather than disappearing. Controlling for participant and exact $(T,\rho)$-cell fixed effects, the adjusted gap falls by $17.6$ points in the High regime (s.e.\ $1.6$; $p<0.001$), while the estimated changes in the Low ($-0.2$ points) and Medium ($+3.0$ points) regimes are close to zero. The High-regime decline appears in all four sessions.

Both the level of overestimation and its incomplete decline parallel \citet{duffyTavits2008}. They report a mean subjective pivotality probability of $33\%$ and an actual frequency of $14.9\%$---an $18.1$-point gap, strikingly close to our $17.6$-point gap. With experience, their subjects move toward the historical frequency of pivotality but generally remain above it; moreover, voting follows subjective beliefs more closely than historical actual frequencies. Our setting adds an important boundary to that learning result. They keep teams fixed partly to facilitate learning and note that random rematching could impede it. Because our subjects are randomly rematched after every round, the High regime decline cannot be within-group learning about a fixed set of partners. Instead, the adjustment persists across new partners and changing values of $T$ and $\rho$. Yet it is neither complete nor uniform across regimes: repeated feedback disciplines pivotality beliefs when realized pivotality is consistently near zero, but does not eliminate overestimation or produce comparable convergence in the Low and Medium regimes.

Systematic overestimation is a level property: it compares mean perceived pivotality with the frequency with which subjects are actually pivotal. It does not reveal whether subjects correctly perceive how a change in the information structure changes their pivotality. We therefore compare the effect of information similarity on perceived pivotality with its effect on actual pivotality. We use pivotality alignment to describe whether the response of ex ante perceived pivotality to information similarity matches the response of the ex post realized frequency of the same pivotal event. The pivotality wedge is perceived minus actual pivotality; the effect of similarity on this wedge indicates whether the two responses move closer together or farther apart.

\begin{result}[Pivotality alignment]\label{res:alignment}
Information similarity changes perceived and actual pivotality in the same direction only in the Medium regime. In the Low regime, it lowers perceived pivotality while raising actual pivotality, producing unrecognized pivotality. In the Medium regime, it raises both, producing aligned pivotality. In the High regime, it raises perceived pivotality without measurably changing actual pivotality, producing illusory pivotality. Unrecognized and illusory pivotality are two forms of response miscalibration.
\end{result}

\begin{table}[!htbp]
\centering
\caption{Information similarity and pivotality alignment}
\label{tab:pivotality-alignment}
\small
\begin{tabular}{@{}llccc@{}}
\toprule
Regime & Pattern & Perceived & Realized & Pivotality \\
 & & pivotality & pivotality & wedge \\
\midrule
Low & Unrecognized pivotality & -6.755*** & +20.226*** & -26.981*** \\
 & & (2.016) & (6.599) & (7.005) \\
\addlinespace
Medium & Aligned pivotality & +4.215** & +11.552** & -7.338 \\
 & & (1.862) & (4.805) & (5.182) \\
\addlinespace
High & Illusory pivotality & +19.188*** & -1.119 & +20.308*** \\
 & & (2.869) & (1.742) & (3.321) \\
\bottomrule
\end{tabular}
\par\vspace{2pt}
\begin{minipage}{0.97\textwidth}
\footnotesize \textit{Notes:} Entries are the estimated percentage-point changes as information similarity $\rho$ moves from 0 to 1. Each column reports a separate regression. Perceived pivotality is the stated probability that exactly $T-1$ other group members report. Realized pivotality is $100\times\mathbf{1}\{A-R_i=T-1\}$. The pivotality wedge is perceived minus realized pivotality. All specifications include exact-threshold, round, and participant fixed effects. Standard errors, two-way clustered by participant and group-round, are in parentheses; $^{*}p<0.1$, $^{**}p<0.05$, $^{***}p<0.01$. The estimation sample consists of 7,035 decisions made after the participant received a warning.
\end{minipage}
\end{table}

The three regimes display distinct response patterns. In Low, perceived pivotality falls by $6.8$ percentage points while actual pivotality rises by $20.2$ points, narrowing the wedge by $27.0$ points: subjects fail to recognize that pivotal opportunities have become more frequent. In Medium, perceived and actual pivotality rise together, and the change in the wedge is not statistically distinguishable from zero. In High, perceived pivotality rises by $19.2$ points while actual pivotality remains essentially unchanged, widening the wedge by $20.3$ points.

These response labels do not replace the level result. Subjects overestimate pivotality in all three regimes: the average perceived--actual gaps are $9.6$ percentage points in Low, $20.9$ points in Medium, and $22.8$ points in High. Static miscalibration is therefore common to all three regimes, whereas response miscalibration takes the distinct forms of unrecognized and illusory pivotality in Low and High.

These response patterns provide a belief-level counterpart to the aggregate outcomes. The next subsection examines whether the reporting induced or discouraged by similarity is converted into content removal and welfare.

\subsection{From Reporting to Removal and Welfare}\label{subsec:removal-welfare}

\begin{result}[Removal tracks reporting only at reachable thresholds]\label{res:removal}
Changes in reporting translate into changes in removal only when the threshold is within reach. Similarity lowers both reporting and harmful-content removal in the Low regime, raises both in the Medium regime, and raises reporting without measurably affecting removal in the High regime.
\end{result}

\begin{figure}[!htbp]
\centering
\includegraphics[width=0.65\textwidth]{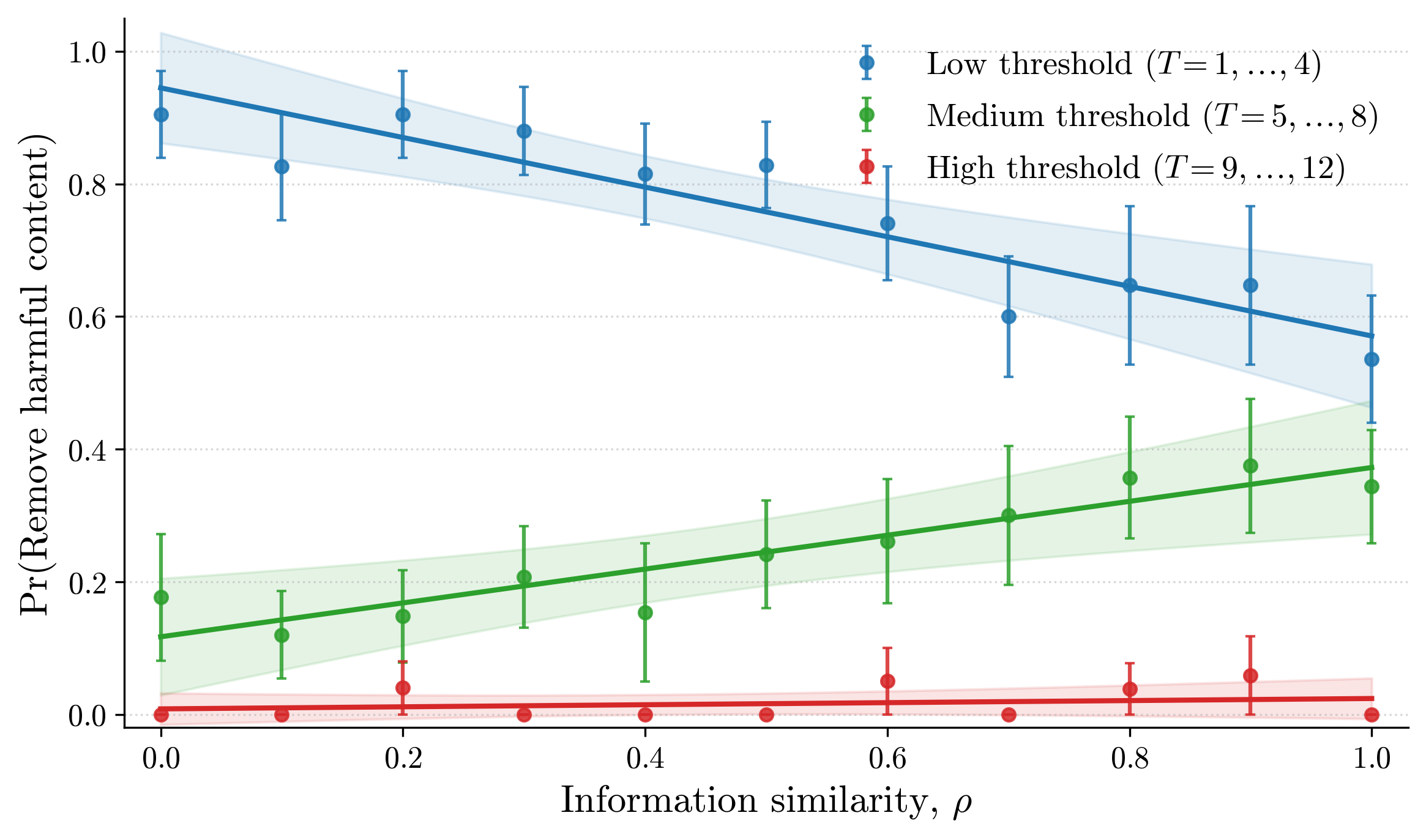}
\caption{Harmful-content removal against information similarity}
\label{fig:group-outcomes}
\caption*{\footnotesize \textit{Notes}: The figure uses harmful group-rounds. Markers show mean removal at each realized value of $\rho$; whiskers are standard errors. Lines are OLS fits with 95 percent confidence bands. The unit of observation is the group-round, and HC1 standard errors are used throughout.}
\end{figure}

Recall from Figure~\ref{fig:report-rho} that similarity lowers individual reporting in the Low regime, raises it in the Medium regime, and raises it most strongly in the High regime. Aggregating these decisions to harmful group-rounds produces the same ordering (Appendix~\ref{app:additional-results}, Table~\ref{tab:hsecondary}, Panel~A): from $\rho=0$ to $\rho=1$, reports fall by about $1.27$ (s.e.\ $0.394$; $p=0.001$) in the Low regime and rise by about $1.30$ (s.e.\ $0.412$; $p=0.002$) and $2.78$ (s.e.\ $0.536$; $p<0.001$) in the Medium and High regimes. Removal, however, does not follow suit. Figure~\ref{fig:group-outcomes} displays the removal patterns, and Panel~B of the same appendix table reports the corresponding regressions. In the Low regime, similarity lowers removal by $0.351$ (s.e.\ $0.090$; $p<0.001$) from a mean of $0.76$; in the Medium regime, it raises removal by $0.246$ (s.e.\ $0.079$; $p=0.002$) from a mean of $0.25$; but in the High regime, removal remains near $0.02$ and does not move ($+0.013$, s.e.\ $0.022$; $p=0.565$)---precisely the regime in which reporting responds most.

With a threshold of nine to twelve in a group of twelve, even a large and coordinated rise in reporting almost never clears the bar. This therefore creates a wedge between effort and success: similarity raises reporting most strongly where it produces almost no additional removal. This wedge between effort and success also determines whether the reporting response creates or destroys welfare, which we examine next.

\begin{result}[Welfare and pivotality alignment]\label{res:welfare}
In our experiment, information similarity lowers welfare under the two response-miscalibrated patterns and raises it under aligned pivotality. Under unrecognized pivotality in Low, welfare falls because reduced reporting prevents groups from reaching an attainable threshold. Under aligned pivotality in Medium, additional reporting increases removal and welfare. Under illusory pivotality in High, additional reporting is costly but rarely changes the collective outcome.
\end{result}

We measure the realized welfare of a group-round as the value of removal to the group net of total reporting costs,
\begin{equation*}\label{eq:welfare}
    W \;=\; \underbrace{\theta\cdot\mathbf{1}\{A\geq T\}\cdot 100\cdot 12}_{\text{benefit}}
    \;-\; \underbrace{30\,A}_{\text{cost}},
\end{equation*}
the $100$-point gain to each of the twelve members when harmful content ($\theta=1$) is removed, less the $30$ points paid by each of the $A$ reporters. We estimate the $\rho$ slope on $W$ from group-round regressions on harmful rounds, with threshold and round fixed effects.

Making information more similar can change welfare for two very different reasons. It can change how a \emph{warned} member behaves---the strategic response documented above---or it can change how \emph{many} members are warned, since a group with more informed members is mechanically more likely to clear the threshold. Only the first is a statement about behavior. Our design lets us hold the second fixed: because CAD keeps each member's marginal warning probability at $0.75$, the ex-ante expected number of warned members ($\approx 9$ on a harmful round) is the same at every $\rho$, so similarity leaves the composition of the informed unchanged on average, and we can read off the behavioral channel by controlling for the realized number of warned members.

Panel~A of Table~\ref{tab:welfarechannels} does exactly this, and the two channels come apart. The number of warned members is a powerful determinant of welfare in its own right---one additional warned member is worth about $+53$, $+32$, and $-7$ points in the Low, Medium, and High regimes (column~3). Yet controlling for it barely moves the similarity slope (columns~1--2), and removal behaves the same way. Similarity therefore changes welfare through the behavior of the informed, not through their number. This is the precise sense in which the welfare result is a \emph{behavioral} one: we strip out the mechanical role of who happens to be informed and isolate what the informed choose to do.

\begin{table}[!htbp]
\centering
\caption{Welfare and removal: behavior, composition, and decomposition.}
\label{tab:welfarechannels}
\small
\begin{tabular}{@{}lccc@{}}
\toprule
\multicolumn{4}{@{}l}{\textit{Panel A. Behavior versus composition}} \\
 & (1) $\rho$ slope & (2) $\rho$ slope & (3) Warned \\
 & baseline & control \#warned & members \\
\midrule
\multicolumn{4}{@{}l}{\textit{Group welfare (points)}} \\
Low & $-383.4^{***}$ & $-383.6^{***}$ & $+53.5^{***}$ \\
 & $(99.1)$ & $(92.2)$ & $(6.9)$ \\
Medium & $+255.8^{***}$ & $+268.5^{***}$ & $+31.7^{***}$ \\
 & $(87.2)$ & $(84.9)$ & $(5.7)$ \\
High & $-67.9^{**}$ & $-66.1^{**}$ & $-7.4^{***}$ \\
 & $(28.5)$ & $(27.1)$ & $(1.7)$ \\
\addlinespace
\multicolumn{4}{@{}l}{\textit{Harmful content removed (0--1)}} \\
Low & $-0.351^{***}$ & $-0.351^{***}$ & $+0.051^{***}$ \\
 & $(0.090)$ & $(0.083)$ & $(0.006)$ \\
Medium & $+0.246^{***}$ & $+0.260^{***}$ & $+0.035^{***}$ \\
 & $(0.079)$ & $(0.075)$ & $(0.005)$ \\
High & $+0.013$ & $+0.012$ & $+0.003^{*}$ \\
 & $(0.022)$ & $(0.022)$ & $(0.002)$ \\
\addlinespace
Threshold, round FE & Yes & Yes & Yes \\
Number-of-warned control & No & Yes & Yes \\
\midrule
\multicolumn{4}{@{}l}{\textit{Panel B. Welfare decomposition controlling for the number warned}} \\
 & (1) Low & (2) Medium & (3) High \\
\midrule
Benefit (removal value) & $-421.6^{***}$ & $+311.9^{***}$ & $+14.5$ \\
 & $(99.7)$ & $(90.4)$ & $(26.4)$ \\
Cost (reporting) & $-38.0^{***}$ & $+43.4^{***}$ & $+80.6^{***}$ \\
 & $(10.8)$ & $(9.1)$ & $(12.0)$ \\
Productive reports & $-0.949^{***}$ & $+1.635^{***}$ & $+0.110$ \\
 & $(0.237)$ & $(0.464)$ & $(0.215)$ \\
Wasted reports & $-0.316$ & $-0.188$ & $+2.578^{***}$ \\
 & $(0.257)$ & $(0.354)$ & $(0.420)$ \\
Observations & 271 & 267 & 242 \\
\bottomrule
\end{tabular}
\par\vspace{2pt}
\begin{minipage}{0.96\textwidth}
\footnotesize \textit{Notes:} All entries come from separate group-round regressions on harmful group-rounds, with threshold and round fixed effects and HC1 standard errors in parentheses. In Panel A, column~(1) reports the similarity slope; column~(2) adds the number of warned members in the group-round; and column~(3) reports that control's coefficient. Panel B reports similarity slopes from the controlled specification, with columns denoting regimes. Benefit equals $\theta\cdot\mathbf{1}\{A\geq T\}\cdot100\cdot12$, cost equals $30A$, and the welfare slope equals the benefit slope minus the cost slope. Productive reports help achieve harmful-content removal (capped at $T$); wasted reports are the remainder. $^{*}p<0.1$, $^{**}p<0.05$, $^{***}p<0.01$.
\end{minipage}
\end{table}

\begin{figure}[!htbp]
\centering
\includegraphics[width=0.7\textwidth]{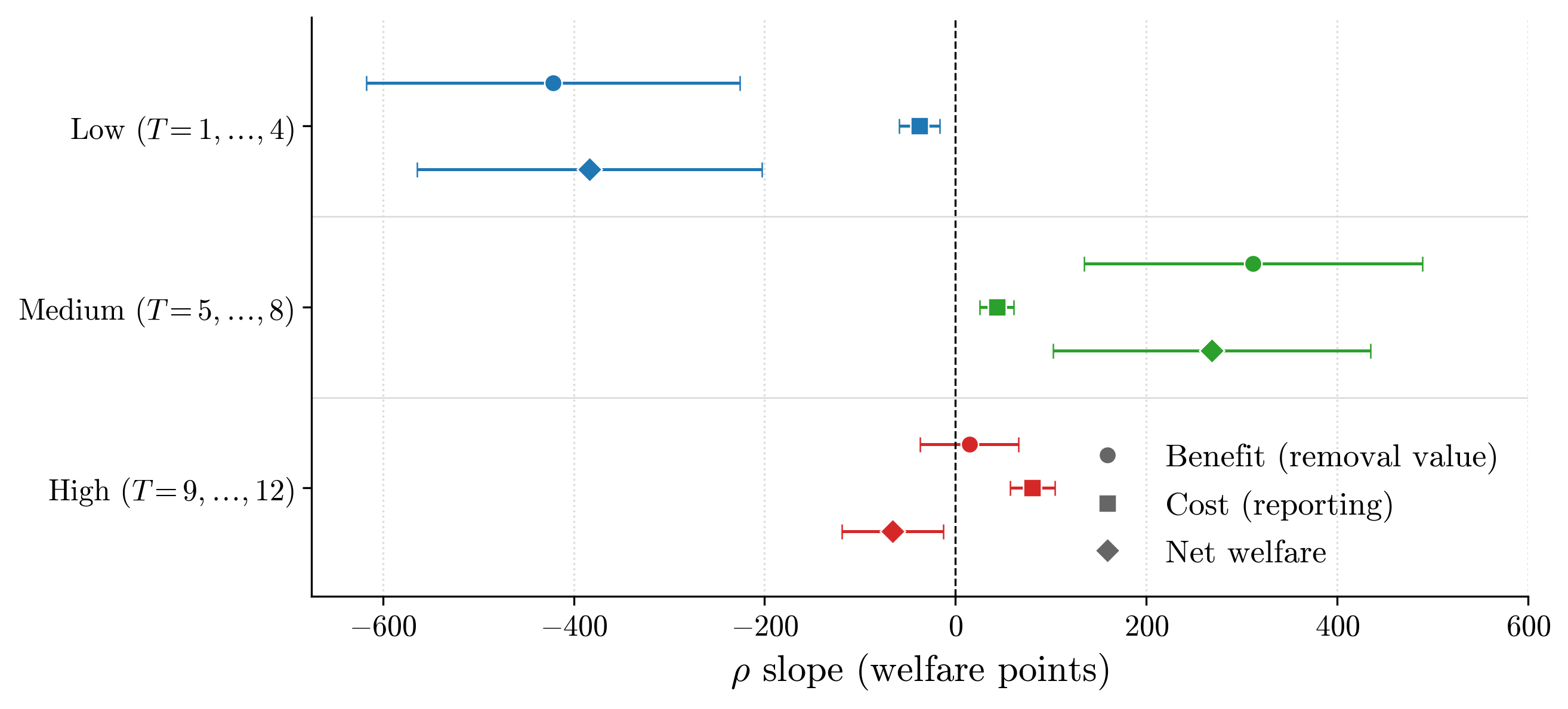}
\caption{Welfare decomposition by regime}
\label{fig:welfare-decomp}
\caption*{\footnotesize \textit{Notes}: Markers are $\rho$ slopes with 95 percent confidence intervals, estimated at the group-round level with threshold and round fixed effects, controlling for the number of warned members in the group-round so the reported slopes hold the composition of the informed fixed (cf.\ Table~\ref{tab:welfarechannels}). Color denotes the regime; circles are the benefit margin, squares the cost margin, and diamonds net welfare. By construction the net welfare slope equals the benefit slope minus the cost slope.}
\end{figure}

Why does the behavioral effect change sign? Because $W$ is linear in its two margins, the $\rho$ slope on welfare splits exactly into a benefit margin ($\theta\cdot\mathbf{1}\{A\geq T\}\cdot100\cdot12$) and a cost margin ($30A$): the welfare slope equals the benefit slope minus the cost slope. We estimate each from a separate group-round regression on harmful rounds, holding the number of warned members fixed, and further split reports into \emph{productive} ones---those that help clear the threshold, capped at $T$---and \emph{wasted} ones. Panel~B of Table~\ref{tab:welfarechannels} and Figure~\ref{fig:welfare-decomp} carry out this split, which runs through a different margin in each regime.

\paragraph{Low threshold: unrecognized pivotality and forgone success.} The Low regime is the aggregate counterpart of unrecognized pivotality. Similarity lowers perceived pivotality and reporting even as actual pivotality becomes more frequent. Welfare falls steeply with similarity, by $383.6$ points (s.e.\ $92.2$; $p<0.001$), almost entirely on the benefit side: the benefit margin falls by $421.6$ points (s.e.\ $99.7$; $p<0.001$), only partly offset by a $38.0$-point saving in cost (s.e.\ $10.8$; $p<0.001$). More similar information leads warned subjects to expect that others will report, so they free-ride; reports fall, and because the threshold is easy to reach in the Low regime, the lost reports reduce removals that would otherwise have occurred. Consistent with this, productive reports fall by $0.949$ (s.e.\ $0.237$; $p<0.001$), while wasted reports do not measurably change ($-0.316$, s.e.\ $0.257$; $p=0.219$). The welfare loss therefore reflects forgone collective success rather than wasted reporting effort: perceived pivotality moves against the pivotal opportunities generated within the group. This is the theory's discouragement case \citep{basakDebKuvalekarCollective}.

\paragraph{Medium threshold: aligned pivotality and productive participation.} The Medium regime is the aligned-pivotality case. Similarity raises both perceived and actual pivotality, and the accompanying increase in reporting helps groups clear a reachable threshold. Welfare rises with similarity, by $268.5$ points (s.e.\ $84.9$; $p=0.002$), again through the benefit margin: benefit rises by $311.9$ points (s.e.\ $90.4$; $p<0.001$), against a $43.4$-point rise in cost (s.e.\ $9.1$; $p<0.001$). Removal increases by $0.260$ (s.e.\ $0.075$; $p<0.001$), and the extra reports are productive ($+1.635$, s.e.\ $0.464$; $p<0.001$) rather than wasted ($-0.188$, s.e.\ $0.354$; $p=0.595$). The threshold in the Medium regime is demanding but reachable, so the coordinated reporting induced by similarity converts into removals. The theory does not sign the welfare effect in this range, but the data identify it as the range in which similar information generates the greatest welfare gain. This is the only regime in our experiment in which the responses of perceived and actual pivotality align and greater participation translates into greater removal and welfare.

\paragraph{High threshold: illusory pivotality and wasted effort.} The High regime is the aggregate counterpart of illusory pivotality. Similarity raises perceived pivotality and reporting, while actual pivotality and content removal remain near zero. Welfare falls with similarity, by $66.1$ points (s.e.\ $27.1$; $p=0.015$). The benefit margin is flat ($+14.5$, s.e.\ $26.4$; $p=0.583$) because removal is stuck near $0.02$, while the cost margin rises by $80.6$ points (s.e.\ $12.0$; $p<0.001$): almost all of the extra reporting is wasted (wasted reports $+2.578$, s.e.\ $0.420$; $p<0.001$; productive reports $+0.110$, s.e.\ $0.215$; $p=0.609$). This is a finite-group boundary condition for the encouragement prediction. Similarity pushes participation in the encouraging direction, but the near-unanimous threshold in the High regime is effectively out of reach: the additional effort rarely clears the threshold, so the benefit channel is nearly closed and welfare falls through higher reporting costs. The resulting participation is therefore largely ineffective: perceived influence expands without a corresponding expansion in actual pivotal opportunities.

Information similarity does what the theory asks of it---it mobilizes participation once the goal is demanding enough that coordination matters---yet mobilization and welfare come apart. The sign of the welfare effect is set by the institution, not by the behavioral response: making information more similar creates value only where the threshold is demanding but reachable, destroys it where the threshold is easy enough to invite free-riding, and merely adds redundant cost where the threshold is out of reach. This tension is most pronounced in the High regime. More similar information has its largest effect on participation exactly where it contributes least to welfare---at the most demanding thresholds, the reporting response is largest and welfare still falls---so it can raise collective effort and lower collective welfare at once. This is a boundary the encouragement logic does not anticipate: in a finite group, mobilizing participation is necessary but not sufficient, because an institution too demanding to convert effort into provision turns the extra effort into pure cost.

\subsection{Heterogeneity}\label{subsec:heterogeneity}

The preceding results document a positive association between perceived pivotality and reporting. We next ask whether this association differs with risk and social preferences. The post-task incentivized games provide four measures: risk tolerance from the multiple price list, the transfer in the dictator game, and the trusting (A2) and trustworthy (B1) choices in the trust game (see Appendix~\ref{app:end_survey}). Appendix~\ref{app:additional-results}, Table~\ref{tab:heterogeneity-descriptives}, reports their distributions. Risk choices are monotonic for 554 of the 576 participants (96.2 percent), who choose the lottery in an average of 4.47 of the nine rows. The remaining 22 participants (3.8 percent) switch back from the sure option to the lottery as the sure amount rises. Because these choices do not identify a unique switch point, the heterogeneity regressions reported below exclude these 22 participants. The average dictator-game transfer is CNY 2.01 (out of CNY 8); 20.3 percent choose A2, and 52.1 percent choose B1 in the trust game.

Preferences could be associated with three distinct margins: the level of reporting, pivotality beliefs, or the strength of the association between them. Appendix~\ref{app:additional-results}, Table~\ref{tab:heterogeneity}, examines the first two, displaying pooled estimates alongside separate estimates for each threshold regime and entering the four trait measures in the same regression. In the pooled sample, a one-standard-deviation increase in altruism is associated with a 4.1-percentage-point increase in reporting (s.e.\ 1.4; $p=0.004$), and subjects who choose B1 report 6.4 points more often (s.e.\ 2.7; $p=0.020$). Risk tolerance instead predicts beliefs: a one-standard-deviation increase is associated with a 2.01-point increase in stated pivotality (s.e.\ 0.72; $p=0.005$). The regime estimates sharpen this pattern. The association between altruism and reporting is largest in the High regime ($+5.9$ points; $p=0.004$), while risk tolerance predicts pivotality beliefs in the Low and Medium regimes ($+2.17$ and $+2.71$ points; $p=0.013$ and $p=0.038$). Altruism also predicts pivotality beliefs in the High regime ($+3.19$ points; $p=0.013$); the remaining associations are less stable across regimes. Although some point estimates vary across regimes, none of the trait--reporting associations displays the statistically supported negative-to-positive reversal that characterizes the similarity effect. Additionally, Appendix~\ref{app:additional-results}, Table~\ref{tab:heterogeneity-crossover}, examines whether controlling for these traits alters the similarity crossover. It does not: the regime-specific similarity slopes are virtually unchanged and remain negative in the Low regime and positive in the Medium and High regimes. Observable differences in risk tolerance, altruism, and trust therefore do not account for the regime-dependent reversal in the effect of similarity.

Appendix~\ref{app:additional-results}, Table~\ref{tab:heterogeneity-translation}, examines the third margin. The positive association between perceived pivotality and reporting remains in all three regimes, and the four interactions between perceived pivotality and the incentivized traits are jointly insignificant ($p=0.322$). Incentivized traits are associated with reporting and subjects' perceived pivotality, but we find no systematic evidence that they alter the association between pivotality beliefs and reporting. The observed pivotality--reporting association therefore appears robust across these dimensions of participant heterogeneity.

\subsection{A Quantal-Response Equilibrium Benchmark}\label{subsec:qre}

The reduced-form estimates establish how reporting changes on average across heterogeneous subjects whose play need not be in equilibrium. They do not by themselves show whether the crossover can arise under equilibrium-consistent play, rather than from aggregating heterogeneous or off-equilibrium responses. We address this concern with the logit quantal-response equilibrium of \citet{mckelveyPalfrey1995}, following the use of QRE by \citet{levinePalfrey2007} to organize noisy participation in laboratory voting games. The benchmark complements the reduced-form evidence.

We compute a type-symmetric QRE of the twelve-person game. A subject's type is her signal, and warned and unwarned subjects report with probabilities $p_W$ and $p_L$. In each exact $(\rho,T)$ cell, these probabilities are logit responses to the expected payoff difference between reporting and ignoring. Pivotality is endogenous: the probability that exactly $T-1$ others report is computed from the same $(p_W,p_L)$ and the signal-correlation structure generated by $\rho$. We estimate one precision parameter $\lambda$ by maximum likelihood on all $11{,}520$ decisions and select the principal QRE branch by continuation from random choice at $\lambda=0$. Appendix~\ref{app:qre} gives the model and estimation details.

The QRE has $\hat\lambda=0.0459$ (s.e.\ $0.0006$) and improves substantially on random choice: its log-likelihood is $-6414.5$, compared with $-7985.1$ when both actions are chosen with equal probability (Table~\ref{tab:qre}). More importantly, it generates the sign change. The predicted similarity slope for warned reporting is negative through $T=5$ and becomes positive at $T=6$, yielding an interpolated crossover of $T^{*}_{\mathrm{QRE}}=5.11$. This is close to the reduced-form estimate of $T^{*}\approx4.5$ in Section~\ref{subsec:reporting-crossover}. The crossover therefore can arise in an internally consistent noisy equilibrium; it does not require aggregation across heterogeneous or off-equilibrium rules.

\begin{figure}[!htbp]
\centering
\includegraphics[width=\textwidth]{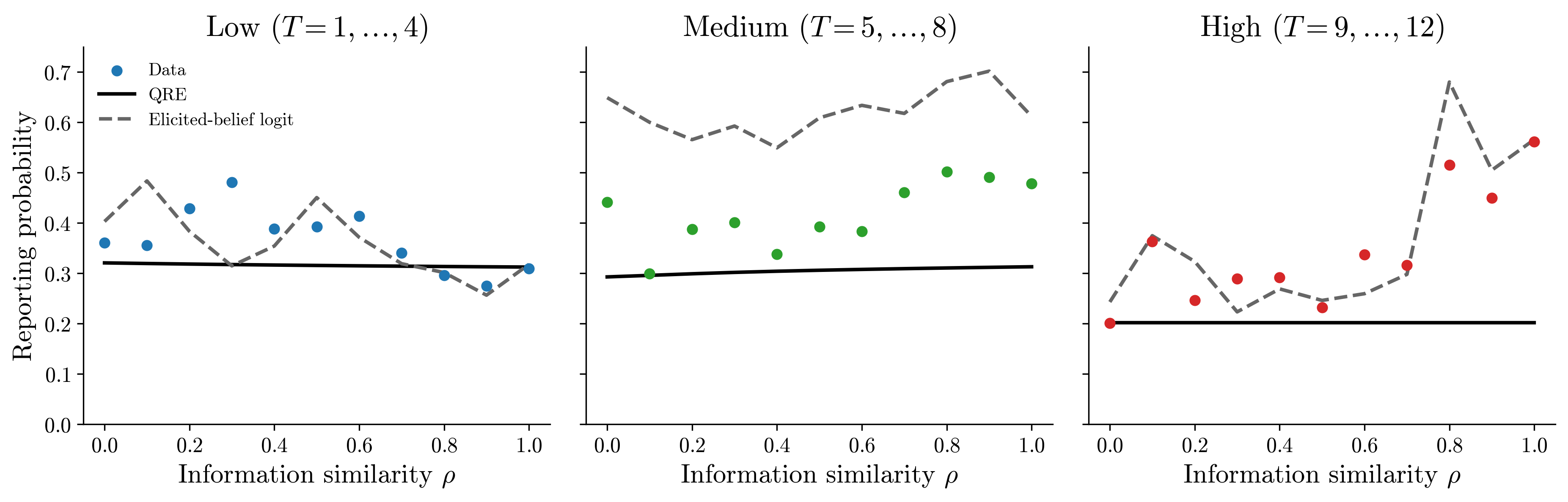}
\caption{Quantal-response benchmark and elicited-belief diagnostic for warned reporting}
\label{fig:qre-regime}
\caption*{\footnotesize \textit{Notes}: Markers show mean reporting at each realized value of $\rho$ among decisions made after the participant received a warning. The solid black line is the logit QRE with equilibrium-consistent pivotality. The dashed line replaces equilibrium pivotality with mean elicited pivotality in each exact $(\rho,T,s)$ cell and re-estimates the logit precision. Model curves give equal weight to the four exact thresholds in each regime. Both precision parameters are estimated on all $11{,}520$ decisions; the elicited-belief logit is not an equilibrium model.}
\end{figure}

The model's largest failure occurs in the High regime's illusory-pivotality case: observed perceived pivotality and reporting rise strongly with similarity even though equilibrium and actual pivotality remain near zero. It captures the direction of the Low-to-Medium reversal but understates its magnitude, and it cannot generate the steep High-regime response (Figure~\ref{fig:qre-regime}, solid line). Once the threshold is near unanimity, equilibrium pivotality is essentially zero. The model therefore predicts a warned reporting rate near its logit floor, about $0.20$, that is almost invariant to $\rho$. Yet Result~\ref{res:calibration} shows that subjects do not hold this model-implied assessment: in the High regime, they assign a $23.3$ percent probability to a pivotal event that occurs in only $0.5$ percent of decisions, and their reporting follows perceived rather than empirical actual pivotality. The QRE thus fails precisely where the distance between equilibrium pivotality and subjects' perceived pivotality is greatest.

To assess whether this documented belief distortion organizes the discrepancy, we replace equilibrium pivotality with the mean elicited pivotality in each exact $(\rho,T,s)$ cell and re-estimate the single precision parameter in the same logit choice rule. The resulting elicited-belief logit is not an equilibrium model, because the elicited beliefs need not be choice-consistent, but like the QRE it contains one estimated choice parameter. Its fitted precision is $\hat\lambda_b=0.0876$ (s.e.\ $0.0017$), and its log-likelihood is $-6065.4$, compared with $-6414.5$ for the QRE. More importantly, it raises predicted High-regime reporting from $0.20$ to $0.37$, close to the observed rate of $0.36$, and restores a strongly positive similarity response (Figure~\ref{fig:qre-regime}, dashed line). It also matches the Low-regime level ($0.36$ against $0.37$) but overshoots the Medium regime ($0.63$ against $0.42$), so it is not a complete account of behavior. Because beliefs are elicited rather than separately assigned, the comparison is diagnostic rather than causal. Its success exactly where the QRE fails nevertheless shows that the High-regime discrepancy is systematically organized by the pivotality overestimation documented in Result~\ref{res:calibration}, rather than by undirected choice noise.

The benchmark therefore delivers two linked findings. QRE reproduces the existence and approximate location of the crossover, showing that the reversal can arise under equilibrium-consistent noisy play. Its High-regime limit is behaviorally structured: when equilibrium pivotality collapses but perceived pivotality remains high, subjects report in line with their perceived rather than actual pivotality. The quantal-response analysis thus corroborates the reduced-form reversal and turns its High-regime failure into evidence about the documented belief miscalibration.

\section{Conclusion}\label{sec:conclusion}

Whether more similar information mobilizes collective action turns on the institution that converts individual actions into collective success. We show this in a twelve-person content-moderation experiment that independently varies information similarity and the removal threshold. Raising $\rho$ increases the probability that group members' algorithmic reports are generated by a single evaluation rather than by separate evaluations, while leaving the marginal informativeness of each subject's own algorithmic report unchanged. The exact threshold varies across rounds within sessions and, through the regime assignment, across sessions, letting us observe how reporting responds as the number of reports required for removal changes. The design therefore isolates the strategic effect of information similarity from both changes in what a subject learns about the content and differences in the participant pool.

The reporting response provides direct support for the theory's central comparative static. Among subjects who received a warning, increasing $\rho$ from zero to one lowers reporting by about 17 percentage points in the Low regime, raises it by about 16 points in the Medium regime, and raises it by about 34 points in the High regime. Pooling across exact thresholds, we estimate that the similarity slope crosses zero at $T^{*}\approx4.5$ in our environment. More similar information therefore discourages reporting when only a few actions are required and encourages it once the collective goal becomes sufficiently demanding. The sign of the effect is not a property of the information structure alone; it depends on whether expected participation by others makes a subject's own report more likely to be pivotal or redundant.

The elicited beliefs provide suggestive evidence that perceived pivotality partially mediates the reporting reversal. As $\rho$ rises, subjects expect more other group members to report in every threshold regime. Perceived pivotality, by contrast, falls in the Low regime and rises in the Medium and High regimes, mirroring the reporting response. Adding this belief to the reporting regression attenuates about one quarter of the High--Low difference in the similarity slope. At the same time, perceived pivotality exceeds its realized frequency by 17.6 percentage points on average, and reporting is positively associated with perceived pivotality. Together, these patterns suggest that reporting is related to whether subjects believe their report can change removal, even though they overestimate how often it actually does.

Calibration in levels and alignment in responses are distinct. Subjects overestimate pivotality in every regime, but perceived and actual pivotality respond to information similarity in the same direction only under intermediate thresholds. Easy thresholds produce unrecognized pivotality, intermediate thresholds aligned pivotality, and near-unanimous thresholds illusory pivotality.

The reporting crossover is robust to observable heterogeneity in risk and social preferences and does not require aggregation of off-equilibrium behavior: a one-parameter logit QRE reproduces its sign change and approximate location. The QRE misses the High-regime response precisely where equilibrium pivotality collapses but subjects continue to overestimate it, linking the model's limit to the belief miscalibration documented above.

These patterns map onto different collective consequences. In our experiment, welfare rises under aligned pivotality. It falls under unrecognized pivotality because consequential participation is discouraged, and under illusory pivotality because costly participation increases without changing collective success.

Two design implications follow. First, information structures and collective-action rules must be evaluated jointly. For content-moderation systems, petitions, whistle-blowing channels, and organizational votes, changing what participants can infer about one another may invite free-riding under one threshold and coordination under another. Second, participation is an intermediate outcome rather than a sufficient measure of institutional performance. A rule can generate more costly action while leaving collective success unchanged, so evaluations that stop at reporting or turnout can reverse the welfare conclusion. The relevant question is not simply whether information similarity mobilizes participants, but whether the institution converts the induced participation into the desired outcome.

Changes in the information structure can therefore move perceived and actual pivotality in opposite directions: whether participation has social value depends on whether perceived influence is aligned with actual pivotal opportunities and whether the institution converts individual actions into collective success.

\newpage
\bibliographystyle{ecca}

\bibliography{references}
\newpage
\appendix
\counterwithin{table}{section}
\counterwithin{figure}{section}

\section{Additional Results}\label{app:additional-results}

This appendix reports the likely-benign sample and additional results supporting Section~\ref{sec:results}: the continuous exact-threshold interaction, the pivotality-belief mediation analysis, the pivotality-calibration comparison, the regression of reporting on perceived and empirical actual pivotality, the group-level reporting and removal regressions, descriptive statistics for the incentivized heterogeneity measures, their associations with reporting and pivotality beliefs, the robustness of the similarity crossover to controlling for these traits, and heterogeneity in the association between pivotality beliefs and reporting.

\paragraph{The likely-benign sample.}
The individual-level analysis uses the warning sample, in which the algorithm reveals that the content is harmful. In the remaining $4{,}485$ decisions the subject instead receives a \textsc{Likely Benign} signal, which leaves substantial doubt about the state. Reporting in this sample is uniformly low. Table~\ref{tab:benign-sample} reports the reporting rate and the slope of reporting on $\rho$ by threshold regime: the rate is $6.35\%$ overall---against $38.4\%$ in the warning sample---and ranges from $4.99\%$ to $7.72\%$ across regimes, with $\rho$ slopes near zero (only the small negative slope in the High regime is distinguishable from zero). Because so few subjects report after a likely-benign signal, this sample carries no test of the strategic mechanism, which concerns how information similarity reshapes reporting \emph{after a warning}; we therefore restrict the individual-level analysis to the warning sample.

\begin{table}[!htbp]
\centering
\caption{Reporting in the likely-benign sample}
\label{tab:benign-sample}
\small
\begin{tabular}{@{}lrcc@{}}
\toprule
 & $N$ & Report rate (\%) & $\rho$ slope \\
\midrule
Low ($T=1$--$4$) & 1{,}413 & 6.72 (1.00) & 0.017 (0.021) \\
Medium ($T=5$--$8$) & 1{,}347 & 7.72 (1.49) & 0.021 (0.026) \\
High ($T=9$--$12$) & 1{,}725 & 4.99 (0.92) & -0.043** (0.018) \\
\midrule
All & 4{,}485 & 6.35 (0.66) & -0.010 (0.012) \\
\bottomrule
\end{tabular}
\par\vspace{2pt}
\begin{minipage}{0.94\textwidth}
\footnotesize \textit{Notes:} The likely-benign sample consists of the 4{,}485 decisions in which the subject received a \textsc{Likely Benign} signal. The reporting rate is the mean of the reporting indicator, in percent; the $\rho$ slope is the coefficient from a linear probability model of reporting on information similarity $\rho$. Standard errors, two-way clustered by participant and group-round, are in parentheses; $^{*}p<0.1$, $^{**}p<0.05$, $^{***}p<0.01$. For comparison, the reporting rate in the warning sample is 38.4 percent.
\end{minipage}
\end{table}

\begin{table}[!htbp]
\centering
\caption{Continuous threshold interaction: the similarity slope as a function of the exact threshold $T$.}
\label{tab:thresholdinteraction}
\small
\begin{tabular}{@{}lcc@{}}
\toprule
 & (1) Report & (2) Pivotality (pp) \\
\midrule
Similarity $\rho$ & -0.253*** & -13.93*** \\
 & (0.043) & (2.63) \\
$\rho\times T$ & 0.056*** & 2.95*** \\
 & (0.006) & (0.41) \\
Constant & 0.324*** & 24.50*** \\
 & (0.011) & (0.70) \\
\midrule
Threshold, round, participant FE & Yes & Yes \\
Observations & 7,035 & 7,035 \\
$R^2$ & 0.435 & 0.433 \\
\bottomrule
\end{tabular}
\par\vspace{2pt}
\begin{minipage}{0.92\textwidth}
\footnotesize \textit{Notes:} Each column regresses the outcome on $\rho$, its interaction with the exact threshold $T$, and threshold, round, and participant fixed effects. The $\rho\times T$ coefficient is the change in the similarity slope for a one-unit increase in the exact threshold. Because each session is assigned to a single threshold regime, the exact threshold varies within sessions across a four-threshold window and between sessions across regimes; the full-range interaction is therefore identified substantially from between-session variation. Standard errors are two-way clustered by participant and group-round in parentheses; $^{*}p<0.1$, $^{**}p<0.05$, $^{***}p<0.01$. The estimation sample consists of decisions made after the participant received a warning.
\end{minipage}
\end{table}

\paragraph{Session-level inference.}
Because each session is assigned to a single threshold regime, the regime contrasts and the exact-threshold interaction draw on variation across the twelve sessions. We therefore re-examine the two headline results of Section~\ref{subsec:reporting-crossover} with inference at the session level: a wild cluster bootstrap that clusters on the twelve sessions (Webb weights, $9{,}999$ replications, imposing the null) and, for the regime contrast, randomization inference that permutes the assignment of sessions to regimes. Table~\ref{tab:session-inference} reports the results. The High-minus-Low reversal in the similarity slope has a wild-cluster-bootstrap $p$-value of $0.001$ and a randomization-inference $p$-value of $0.0001$, and the $\rho\times T$ interaction has a bootstrap $p$-value of $0.001$. The reversal is also visible session by session: each of the four High-regime sessions has a more positive similarity slope than each of the four Low-regime sessions.

\begin{table}[!htbp]
\centering
\caption{Session-level inference for the reporting crossover}
\label{tab:session-inference}
\small
\begin{tabular}{@{}lccc@{}}
\toprule
 & Estimate & Wild cluster & Randomization \\
 & & bootstrap $p$ & inference $p$ \\
\midrule
High$-$Low similarity slope & 0.509 & 0.0007 & 0.0001 \\
Similarity $\times$ exact threshold ($\rho\times T$) & 0.056 & 0.0008 & --- \\
\bottomrule
\end{tabular}
\par\vspace{2pt}
\begin{minipage}{0.94\textwidth}
\footnotesize \textit{Notes:} Because each of the twelve sessions is assigned to a single threshold regime (four sessions per regime), the between-regime comparisons are also between-session comparisons. The two headline results of Section~\ref{subsec:reporting-crossover}---the High-minus-Low difference in the similarity slope and the $\rho\times T$ interaction---are re-examined with inference at the session level. The wild cluster bootstrap clusters on the twelve sessions, imposes the null, and uses Webb six-point weights with $9{,}999$ replications, partialling out the participant, threshold, and round fixed effects. Randomization inference enumerates all $34{,}650$ assignments of the twelve sessions to three regimes of four and compares the observed High-minus-Low gap in the session-level similarity slopes with the permutation distribution. This permutation tests the between-regime contrast; the $\rho\times T$ interaction is a continuous slope in the exact threshold rather than a between-regime contrast, so for it we report only the wild cluster bootstrap. Both results remain significant at the session level.
\end{minipage}
\end{table}

\begin{table}[!htbp]
\centering
\caption{Pivotality beliefs partially mediate the High$-$Low similarity gap in reporting.}
\label{tab:hmediation}
\small
\begin{tabular}{@{}lcc@{}}
\toprule
 & (1) Baseline & (2) With pivotality belief \\
\midrule
High $-$ Low similarity slope ($\rho$) & 0.509*** & 0.385*** \\
 & (0.054) & (0.048) \\
\addlinespace
Pivotality belief (per pp) &  & 0.0048*** \\
 &  & (0.0003) \\
\midrule
Implied attenuation &  & 24.3\% \\
Observations & 7,035 & 7,035 \\
$R^2$ & 0.437 & 0.479 \\
\bottomrule
\end{tabular}
\par\vspace{2pt}
\begin{minipage}{0.92\textwidth}
\footnotesize \textit{Notes:} The dependent variable is the reporting indicator. Both columns are linear probability models with threshold, participant, and round fixed effects and the full set of similarity main and interaction terms; only the High$-$Low similarity interaction is reported. Column (2) adds the subject's elicited probability that exactly $T-1$ other group members report, measured in percentage points. Attenuation is $1-\beta^{(2)}/\beta^{(1)}$ for the High$-$Low interaction. Standard errors are two-way clustered by participant and group-round in parentheses; $^{*}p<0.1$, $^{**}p<0.05$, $^{***}p<0.01$. The estimation sample consists of decisions made after the participant received a warning.
\end{minipage}
\end{table}

\begin{table}[!htbp]
\centering
\caption{Perceived and actual pivotality}
\label{tab:pivotality-calibration}
\small
\begin{tabular}{lrrrrr}
\toprule
 & & Perceived & Actual & \multicolumn{2}{c}{Gap} \\
\cmidrule(lr){5-6}
Sample & $N$ & (\%) & (\%) & (pp) & s.e. \\
\midrule
All & 7,035 & 27.7 & 10.2 & +17.6*** & (1.1) \\
Low regime & 2,427 & 22.9 & 13.3 & +9.6*** & (2.1) \\
Medium regime & 2,493 & 36.1 & 15.2 & +20.9*** & (1.9) \\
High regime & 2,115 & 23.3 & 0.5 & +22.8*** & (1.5) \\
\bottomrule
\end{tabular}
\par\vspace{2pt}
\begin{minipage}{0.94\textwidth}
\footnotesize \textit{Notes:} Let $A$ denote total reports in the group and $R_i$ subject $i$'s reporting decision. Actual pivotality is the realized indicator $\mathbf{1}\{A-R_i=T-1\}$: it equals one exactly when changing subject $i$'s action would change removal. Perceived pivotality is the elicited probability of this same event, and the gap is perceived minus actual pivotality. Entries are observation-level means. Standard errors for the mean gap are two-way clustered by participant and group-round. $^{*}p<0.1$, $^{**}p<0.05$, $^{***}p<0.01$. The estimation sample consists of decisions made after the participant received a warning.
\end{minipage}
\end{table}

\begin{table}[!htbp]
\centering
\caption{Perceived and empirical actual pivotality as predictors of reporting}
\label{tab:calibration-reporting}
\small
\begin{tabular}{@{}lc@{}}
\toprule
 & (1) Report \\
\midrule
Perceived pivotality (10 pp) & 0.047*** \\
 & (0.003) \\
\addlinespace
Empirical actual-pivotality benchmark (10 pp) & 0.002 \\
 & (0.005) \\
\midrule
Similarity terms & Yes \\
Threshold, round, participant FE & Yes \\
Observations & 6,967 \\
$R^2$ & 0.479 \\
\bottomrule
\end{tabular}
\par\vspace{2pt}
\begin{minipage}{0.92\textwidth}
\footnotesize \textit{Notes:} The dependent variable is the reporting indicator. The table reports a linear probability model that includes perceived pivotality and an empirical actual-pivotality benchmark alongside the full set of similarity main and regime-interaction terms and threshold, round, and participant fixed effects. Perceived pivotality is the subject's stated probability that exactly $T-1$ other group members report. The empirical benchmark is the realized pivotality frequency in other group-rounds with the same exact $(T,\rho)$, excluding the subject's current group-round. Coefficients give the change in reporting probability associated with a 10-percentage-point increase in each measure. Standard errors are two-way clustered by participant and group-round in parentheses; $^{*}p<0.1$, $^{**}p<0.05$, $^{***}p<0.01$. The estimation sample consists of decisions made after the participant received a warning.
\end{minipage}
\end{table}

\begin{table}[!htbp]
\centering
\caption{Information similarity, aggregate reporting, and harmful-content removal.}
\label{tab:hsecondary}
\small
\begin{tabular}{@{}lccc@{}}
\toprule
 & (1) Low & (2) Medium & (3) High \\
 & ($T=1$--$4$) & ($T=5$--$8$) & ($T=9$--$12$) \\
\midrule
\multicolumn{4}{@{}l}{\textit{Panel A. Number of reports}} \\
Similarity $\rho$ & $-1.265^{***}$ & $1.301^{***}$ & $2.775^{***}$ \\
 & $(0.394)$ & $(0.412)$ & $(0.536)$ \\
Constant & $4.135^{***}$ & $3.350^{***}$ & $1.795^{***}$ \\
 & $(0.225)$ & $(0.224)$ & $(0.265)$ \\
$R^2$ & 0.24 & 0.23 & 0.37 \\
\addlinespace
\multicolumn{4}{@{}l}{\textit{Panel B. Harmful content removed}} \\
Similarity $\rho$ & $-0.351^{***}$ & $0.246^{***}$ & $0.013$ \\
 & $(0.090)$ & $(0.079)$ & $(0.022)$ \\
Constant & $0.934^{***}$ & $0.106^{**}$ & $0.009$ \\
 & $(0.047)$ & $(0.043)$ & $(0.012)$ \\
$R^2$ & 0.14 & 0.23 & 0.13 \\
\midrule
Threshold, round FE & Yes & Yes & Yes \\
Observations & 271 & 267 & 242 \\
\bottomrule
\end{tabular}
\par\vspace{2pt}
\begin{minipage}{0.92\textwidth}
\footnotesize \textit{Notes:} Each column within a panel is a separate group-round regression of the panel outcome on similarity $\rho$, with threshold and round fixed effects. The dependent variable in Panel~A is the total number of reports in the group; the dependent variable in Panel~B is a 0--1 indicator for harmful-content removal. The estimation sample in each column consists of harmful group-rounds. HC1 standard errors are in parentheses; $^{*}p<0.1$, $^{**}p<0.05$, $^{***}p<0.01$.
\end{minipage}
\end{table}

\begin{table}[!htbp]
\centering
\caption{Descriptive statistics for incentivized traits}
\label{tab:heterogeneity-descriptives}
\small
\begin{tabular}{@{}lrrrrrr@{}}
\toprule
 & Mean & SD & Median & Min. & Max. & $N$ \\
\midrule
Number of lottery choices (monotonic respondents) & 4.466 & 1.458 & 4.0 & 0.0 & 9.0 & 554 \\
Altruism: transfer (CNY) & 2.009 & 1.761 & 2.0 & 0.0 & 8.0 & 576 \\
Trusting choice (A2) & 0.203 & 0.403 & 0.0 & 0.0 & 1.0 & 576 \\
Trustworthy choice (B1) & 0.521 & 0.500 & 1.0 & 0.0 & 1.0 & 576 \\
\bottomrule
\end{tabular}
\par\vspace{2pt}
\begin{minipage}{0.94\textwidth}
\footnotesize \textit{Notes:} All measures come from incentivized games administered after the main task. Risk tolerance is the number of lottery choices before the first sure choice among respondents with monotonic multiple-price-list choices; higher values indicate greater risk tolerance. Twenty-two participants (3.8 percent) switched back from the sure option to the lottery as the sure amount rose; they are excluded from this row and the heterogeneity regressions. Altruism is the amount allocated to the other participant in the dictator game. A2 is the trusting choice and B1 the trustworthy choice in the trust game.
\end{minipage}
\end{table}

\begin{table}[!htbp]
\centering
\caption{Incentivized traits, reporting, and pivotality beliefs}
\label{tab:heterogeneity}
\footnotesize
\begin{tabular}{@{}lcccc@{}}
\toprule
 & (1) Pooled & (2) Low & (3) Medium & (4) High \\
\midrule
\multicolumn{5}{@{}l}{\textit{Panel A. Reporting}} \\
Risk tolerance (per SD) & 0.009 & -0.001 & -0.005 & 0.040* \\
 & (0.013) & (0.021) & (0.025) & (0.022) \\
Altruism (per SD) & 0.041*** & 0.048* & 0.024 & 0.059*** \\
 & (0.014) & (0.026) & (0.026) & (0.021) \\
Trusting choice (A2) & 0.035 & -0.028 & 0.040 & 0.124** \\
 & (0.035) & (0.058) & (0.064) & (0.055) \\
Trustworthy choice (B1) & 0.064** & 0.042 & 0.085* & 0.055 \\
 & (0.027) & (0.052) & (0.051) & (0.039) \\
\addlinespace[2pt]
$R^2$ & 0.071 & 0.060 & 0.045 & 0.183 \\
\midrule
\multicolumn{5}{@{}l}{\textit{Panel B. Pivotality probability (pp)}} \\
Risk tolerance (per SD) & 2.011*** & 2.171** & 2.708** & 1.094 \\
 & (0.717) & (0.878) & (1.302) & (1.453) \\
Altruism (per SD) & 1.072 & 1.184 & -0.376 & 3.186** \\
 & (0.686) & (1.081) & (1.176) & (1.286) \\
Trusting choice (A2) & -0.270 & -3.059 & 5.569 & -2.858 \\
 & (1.871) & (2.728) & (3.503) & (3.275) \\
Trustworthy choice (B1) & 1.481 & -1.583 & 3.752 & 3.158 \\
 & (1.399) & (2.248) & (2.701) & (2.256) \\
\addlinespace[2pt]
$R^2$ & 0.111 & 0.094 & 0.042 & 0.197 \\
\midrule
Similarity controls & Yes & Yes & Yes & Yes \\
Threshold and round FE & Yes & Yes & Yes & Yes \\
Observations & 6,751 & 2,388 & 2,325 & 2,038 \\
\bottomrule
\end{tabular}
\par\vspace{2pt}
\begin{minipage}{0.96\textwidth}
\scriptsize \textit{Notes:} Columns (1)--(4) report the pooled, Low-, Medium-, and High-regime estimates. Panel A is a linear probability model for reporting; Panel B uses the elicited probability that exactly $T-1$ other group members report. The four trait measures enter each regression simultaneously, so each coefficient holds the other three constant. The pooled specifications allow the coefficient on $\rho$ to differ by regime; the regime-specific specifications control for $\rho$. All specifications include exact-threshold and round fixed effects. Risk tolerance and altruism are standardized; the trust-game indicators are demeaned. Standard errors are two-way clustered by participant and group-round in parentheses; $^{*}p<0.1$, $^{**}p<0.05$, $^{***}p<0.01$. The estimation sample consists of decisions made after a warning by the 554 participants with monotonic risk choices.
\end{minipage}
\end{table}

\begin{table}[!htbp]
\centering
\caption{Similarity crossover with incentivized-trait controls}
\label{tab:heterogeneity-crossover}
\footnotesize
\begin{tabular}{@{}lcc@{}}
\toprule
 & (1) No trait controls & (2) Trait controls \\
\midrule
Similarity slope, Low & -0.161*** & -0.161*** \\
 & (0.040) & (0.040) \\
Similarity slope, Medium & 0.152*** & 0.155*** \\
 & (0.031) & (0.031) \\
Similarity slope, High & 0.314*** & 0.323*** \\
 & (0.044) & (0.043) \\
\midrule
Trait controls & No & Yes \\
Threshold and round FE & Yes & Yes \\
Observations & 6,751 & 6,751 \\
$R^2$ & 0.052 & 0.071 \\
\bottomrule
\end{tabular}
\par\vspace{2pt}
\begin{minipage}{0.88\textwidth}
\footnotesize \textit{Notes:} Both columns report pooled linear probability models of reporting and allow the coefficient on $\rho$ to differ by regime. The displayed coefficients are the implied similarity slopes in each regime. Column (2) adds risk tolerance, altruism, and the trusting (A2) and trustworthy (B1) choices as controls. Risk tolerance and altruism are standardized; the trust-game indicators are demeaned. Both specifications include exact-threshold and round fixed effects. Standard errors are two-way clustered by participant and group-round in parentheses; $^{*}p<0.1$, $^{**}p<0.05$, $^{***}p<0.01$. The estimation sample consists of 6,751 decisions made after a warning by the 554 participants with monotonic risk choices.
\end{minipage}
\end{table}

\begin{table}[!htbp]
\centering
\caption{Incentivized traits and the association between pivotality beliefs and reporting}
\label{tab:heterogeneity-translation}
\footnotesize
\begin{tabular}{@{}lc@{}}
\toprule
 & (1) Report \\
\midrule
Pivotality belief (10 pp), Low & 0.018*** \\
 & (0.006) \\
Pivotality belief (10 pp), Medium & 0.039*** \\
 & (0.005) \\
Pivotality belief (10 pp), High & 0.079*** \\
 & (0.004) \\
\addlinespace
Risk tolerance (per SD) $\times$ pivotality belief & 0.003 \\
 & (0.003) \\
Altruism (per SD) $\times$ pivotality belief & -0.006* \\
 & (0.003) \\
Trusting choice (A2) $\times$ pivotality belief & 0.004 \\
 & (0.007) \\
Trustworthy choice (B1) $\times$ pivotality belief & 0.004 \\
 & (0.006) \\
\addlinespace
Trait-by-pivotality terms (joint $p$-value) & 0.322 \\
\midrule
Similarity controls & Yes \\
Participant, threshold, and round FE & Yes \\
Observations & 6,751 \\
$R^2$ & 0.497 \\
\bottomrule
\end{tabular}
\par\vspace{2pt}
\begin{minipage}{0.88\textwidth}
\footnotesize \textit{Notes:} The table reports a pooled linear probability model of reporting. The first three coefficients allow the association between perceived pivotality and reporting to differ by regime. Each interaction coefficient shows how that association varies with the corresponding trait. The coefficient on $\rho$ is also allowed to differ by regime. The model includes participant, exact-threshold, and round fixed effects. Risk tolerance and altruism are standardized; the trust-game indicators are demeaned. Standard errors are two-way clustered by participant and group-round in parentheses; $^{*}p<0.1$, $^{**}p<0.05$, $^{***}p<0.01$. Significance stars use unadjusted $p$-values. The estimation sample consists of 6,751 decisions made after a warning by the 554 participants with monotonic risk choices.
\end{minipage}
\end{table}

\clearpage

\section{Quantal-Response Equilibrium: Model and Estimation}\label{app:qre}

\paragraph{Signal types and equilibrium pivotality.}
A subject's type is her signal $s\in\{W,L\}$, where $W$ denotes a warning and $L$ a likely-benign signal. A warning reveals harmful content, so $\Pr(\theta=1\mid W)=1$. Under the harmful-content prior of $0.8$, a likely-benign signal implies $\Pr(\theta=1\mid L)=0.5$. Let $p_W$ and $p_L$ denote the type-symmetric probabilities of reporting. Conditional on harmful content, a subject in the independent-evaluation branch expects another member to report with probability
\[
q=0.75p_W+0.25p_L.
\]
The probability that a type-$s$ subject is pivotal is therefore
\begin{equation}\label{eq:qre-pivotality}
\pi_s(\rho,T)
=
\rho\,\mathrm{Bin}\!\left(T-1;11,p_s\right)
+(1-\rho)\,\mathrm{Bin}\!\left(T-1;11,q\right),
\qquad s\in\{W,L\},
\end{equation}
where $\mathrm{Bin}(k;n,p)$ is the probability of exactly $k$ successes in $n$ independent Bernoulli trials. In the common-evaluation branch, all eleven other members share the focal subject's signal but make independent quantal choices; in the independent-evaluation branch, each other member receives a warning with probability $0.75$ conditional on harmful content.

\paragraph{Logit QRE.}
The expected payoff differences between reporting and ignoring are
\[
\Delta_W=100\pi_W-30,
\qquad
\Delta_L=100(0.5)\pi_L-30.
\]
The type-symmetric logit QRE in each exact $(\rho,T)$ cell solves
\begin{equation}\label{eq:qre-fixed-point}
p_s=\Lambda(\lambda\Delta_s)
=\frac{1}{1+\exp(-\lambda\Delta_s)},
\qquad s\in\{W,L\},
\end{equation}
where $\lambda\geq0$ is the precision of the quantal response. At $\lambda=0$, each action is chosen with probability one half; higher $\lambda$ makes choices more responsive to expected-payoff differences. Because $\pi_s$ in equation~\eqref{eq:qre-pivotality} depends on $(p_W,p_L)$, equation~\eqref{eq:qre-fixed-point} is a two-dimensional fixed point.

We trace the principal branch from $(p_W,p_L)=(0.5,0.5)$ at $\lambda=0$ by homotopy continuation and estimate a single $\lambda$ by maximum likelihood on all $11{,}520$ decisions. The reported standard error uses the inverse observed information. At the fitted value, solving directly from initial probabilities $(0.01,0.01)$, $(0.5,0.5)$, and $(0.99,0.99)$ converges to the principal-branch solution in every $(\rho,T)$ cell. To locate the QRE crossover, we regress the fitted $p_W$ on the eleven experimental values of $\rho$ separately at each exact threshold and linearly interpolate between the last negative and first positive slope. The QRE column of Table~\ref{tab:qre} reports the resulting precision estimate, model fit, crossover, and regime-level predictions.

\paragraph{Elicited-belief diagnostic.}
The diagnostic replaces equilibrium pivotality with elicited pivotality while keeping the logit response rule. Let $\bar b_{\rho Ts}$ be the mean stated probability of exactly $T-1$ other reports in the exact $(\rho,T,s)$ cell. The resulting expected-payoff index is
\[
\widetilde\Delta_{\rho Ts}
=100\,\Pr(\theta=1\mid s)\bar b_{\rho Ts}-30.
\]
The elicited-belief logit is
\[
\widetilde p_{\rho Ts}
=\Lambda\!\left(\lambda_b\widetilde\Delta_{\rho Ts}\right).
\]
We estimate $\lambda_b$ by maximum likelihood on all decisions. Because elicited beliefs need not equal the pivotality implied by the fitted reporting probabilities, this specification is not an equilibrium model. It is a one-parameter diagnostic of how well the pivotality beliefs documented in Result~\ref{res:calibration} organize the behavior that the QRE leaves unexplained. Table~\ref{tab:qre} reports both specifications.

\begin{table}[!htbp]
\centering
\caption{Quantal-response benchmark}
\label{tab:qre}
\footnotesize
\begin{tabular}{@{}lcc@{}}
\toprule
 & QRE & Elicited-belief logit \\
\midrule
Precision & $\lambda=0.0459$ & $\lambda_b=0.0876$ \\
 & $(0.0006)$ & $(0.0017)$ \\
Log-likelihood & $-6414.5$ & $-6065.4$ \\
Crossover $T^{*}_{\mathrm{QRE}}$ & $5.1$ & --- \\
\addlinespace
\multicolumn{3}{@{}l}{\textit{Predicted vs.\ actual reporting after a warning}} \\
Low regime (actual $0.37$) & $0.32$ & $0.36$ \\
Medium regime (actual $0.42$) & $0.31$ & $0.63$ \\
High regime (actual $0.36$) & $0.20$ & $0.37$ \\
\bottomrule
\end{tabular}
\par\vspace{2pt}
\begin{minipage}{0.9\textwidth}
\footnotesize \textit{Notes:} Logit quantal-response equilibrium of the twelve-person content-moderation game, estimated by maximum likelihood on all $11{,}520$ decisions. The QRE uses equilibrium pivotality. The elicited-belief logit replaces it with mean elicited pivotality in each exact $(\rho,T,s)$ cell and re-estimates a single precision parameter. It is not an equilibrium model because elicited beliefs need not be choice-consistent. Inverse-observed-information standard errors for estimated precision parameters are in parentheses. $T^{*}_{\mathrm{QRE}}$ is the threshold at which the QRE similarity slope turns positive. Predicted and actual reporting rates use decisions made after the participant received a warning. The random-choice log-likelihood is $-7985.1$.
\end{minipage}
\end{table}

\clearpage
\begin{figure}[!htbp]
\centering
\includegraphics[width=0.8\textwidth]{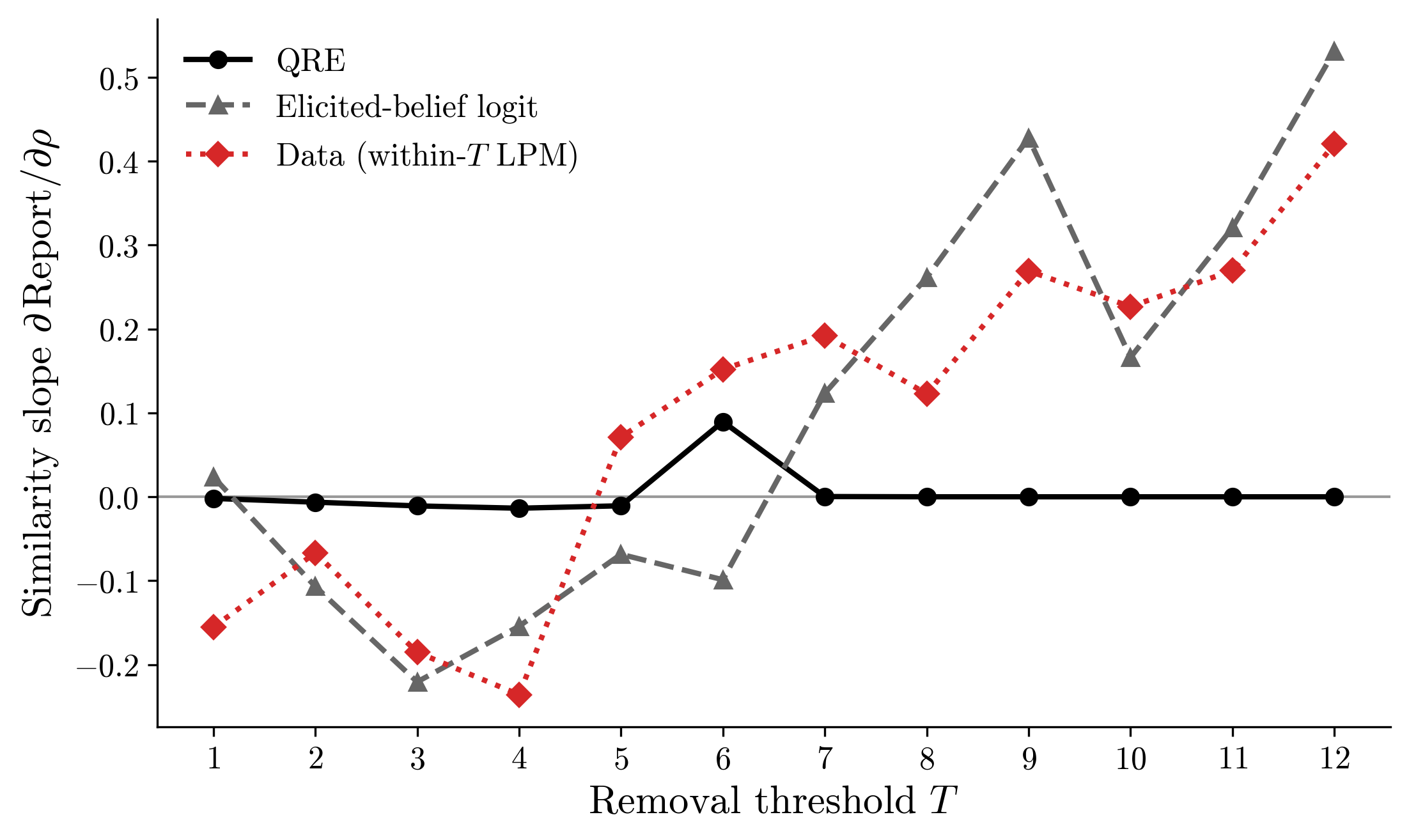}
\caption{Similarity slopes in the QRE benchmark and elicited-belief diagnostic}
\label{fig:qre-slope}
\caption*{\footnotesize \textit{Notes}: The solid black series plots the slope from regressing the fitted QRE warning-type reporting probability on $\rho$ separately at each exact threshold. The dashed series applies the same calculation to the elicited-belief logit. The dotted red series reports the coefficient from a separate linear probability model of reporting on $\rho$ within each exact threshold; these regressions use decisions made after the participant received a warning.}
\end{figure}

\clearpage

\section{Experimental Materials}\label{app:materials}

\subsection{Instructions}\label{app:instructions}

\textit{Paper appendix note: Bracketed labels identify treatment-specific wording. Participants receive only the version assigned to their session and do not see the bracketed labels or the alternative versions.}

Welcome! You are about to participate in a session on decision-making, and you will be paid for your participation with cash, privately at the end of the session. The entire session will take place through computer terminals and there will be no interaction with participants seated at other terminals. Please make sure that you have turned in your cell phone and any other electronic communication devices. Using such devices during the experiment is not allowed.

Please read these instructions carefully. Your earnings will depend on your decisions, the decisions of other participants, and random events determined by the computer. If anything is unclear, please raise your hand and an experimenter will assist you.

You will have 20 minutes to read the printed instructions and complete the comprehension quiz. There is no need to rush. The main task will begin only after all participants have completed the quiz, so finishing early means that you will wait for others. Please use the time to read the instructions carefully. All answers to the comprehension quiz can be found in these instructions. You must answer all quiz questions correctly before you can continue to the experiment. If you answer any question incorrectly, you will need to answer the quiz again until all questions are answered correctly. If you answer all quiz questions correctly on your first attempt, you will receive an additional bonus of \textbf{CNY 5}.

In the main task described in these instructions, earnings are denominated in \textbf{points}. Points earned are converted into cash at the rate \textbf{100 points = CNY 50}. For the show-up fee, main task, and bonus questions described in these instructions:
\begin{enumerate}
    \item you receive a show-up fee of \textbf{CNY 20};
    \item one round of the main task is randomly selected for payment at the end of the experiment;
    \item in the round that is randomly selected for payment, the maximum money to be earned is \textbf{CNY 50};
    \item in addition to the round selected, there will be two bonus questions where you can earn a maximum of \textbf{CNY 20};
    \item if you answer all comprehension quiz questions correctly on your first attempt, you receive an additional bonus of \textbf{CNY 5}.
\end{enumerate}
This means that for the show-up fee, main task, bonus questions, and comprehension-quiz bonus described in these instructions, the maximum payment from this part of the experiment is \textbf{CNY 95}. After the main task, you will complete an end survey. Some parts of the end survey are also incentivized and can generate additional earnings. The detailed rules for those parts will be explained when the end survey begins. Your total payment will include this part of the experiment plus any additional earnings from the incentivized parts of the end survey, and your final payment will always be positive.

All decisions are anonymous. No participant will learn the identity of any other participant, and no participant will be informed of your individual choices.

\paragraph{Description of the Main Task}

In each round, you will be placed in a group of \textbf{12 participants}. Your group members change from round to round. Thus, each round should be treated as a separate decision problem.

In every round, your group considers one piece of online content from a certain platform. The content is either \textbf{Harmful} or \textbf{Benign}. Your task is to decide whether to \textbf{report} the content or \textbf{ignore} the content. If enough participants in your group report the content, the content is removed. Whether removal is desirable depends on whether the content is actually harmful.

At the beginning of each round, the computer randomly picks one piece of content from this platform. It is known that: with probability 0.8, the content is \textbf{Harmful}; with probability 0.2, the content is \textbf{Benign}. You will \textbf{not} know the true content type when you make your decision. The true type is revealed only after all decisions have been made.

To aid your assessment, an ALGORITHM generates an evaluation report of the content:

\begin{itemize}
\item If the content is \textbf{Benign}, the ALGORITHM's evaluation report is \textsc{Likely Benign} with 100\% probability.
\item If the content is \textbf{Harmful}, the ALGORITHM's evaluation report is \textsc{Likely Harmful} with 75\% probability and \textsc{Likely Benign} with 25\% probability.
\end{itemize}

Thus, observing \textsc{Likely Harmful} is strong evidence that the content is \textbf{Harmful}. Observing \textsc{Likely Benign} does not imply that the content is definitely \textbf{Benign}.

However, the ALGORITHM does not always generate evaluation reports independently for each participant:
\begin{itemize}
    \item With probability $\rho$ (which will be shown on your screen in each round), the ALGORITHM performs a single evaluation and sends the same evaluation report to all 12 participants in the group.
    \item With probability $1 - \rho$, the ALGORITHM performs independent evaluations for each participant and sends independent evaluation reports to each individual (you will only observe your own evaluation report).
\end{itemize}

Thus, your own evaluation report is equally accurate in every round: the probability that it shows \textsc{Likely Harmful} when the content is \textbf{Harmful} is always 75\%. The only thing that changes is whether all 12 participants in your group receive the same evaluation report or receive independent evaluation reports.

\paragraph{Action}

After observing the evaluation report and resulting generation process, you choose one of two actions: \textbf{Report} or \textbf{Ignore}.

\begin{center}
\begin{tikzpicture}[
  every node/.style={font=\normalsize},
  arrow/.style={-{Stealth[length=6pt]}, thick}
]
\node (action) at (0,0) {\textbf{Action}};
\node (report) at (3.5, 1.2) {\textbf{Report}};
\node (reportcost) at (6.2, 1.2) {(Cost 30 Points)};
\node (ignore) at (3.5, -1.2) {\textbf{Ignore}};
\node (ignorecost) at (6, -1.2) {(No Cost)};
\draw[arrow] (action) -- (report);
\draw[arrow] (action) -- (ignore);
\end{tikzpicture}
\end{center}

If you choose to \textbf{Report}, you incur a cost of 30 points regardless of whether the content is removed and regardless of whether the content is harmful or benign. If you choose to \textbf{Ignore}, you incur no cost.

\paragraph{Outcome (Threshold $T$)}

Let the total number of participants in your 12-person group who choose to \textbf{Report} be denoted by $A$. The content is removed if and only if the number of participants who choose to \textbf{Report} reaches or exceeds a threshold $T$:
\begin{equation*}
    A \geq T.
\end{equation*}
The value of $T$ will be shown on your screen in each round.

\begin{quote}
\textit{[High-threshold version:]} In each round, $T$ will be one of 9, 10, 11, or 12. The exact value of $T$ will be shown on your screen before you make your decision.

\textit{[Medium-threshold version:]} In each round, $T$ will be one of 5, 6, 7, or 8. The exact value of $T$ will be shown on your screen before you make your decision.

\textit{[Low-threshold version:]} In each round, $T$ will be one of 1, 2, 3, or 4. The exact value of $T$ will be shown on your screen before you make your decision.
\end{quote}

\paragraph{Payoffs}

Your payoff in a round depends on:
\begin{enumerate}
    \item whether you chose to \textbf{Report} or to \textbf{Ignore};
    \item whether the content was removed;
    \item whether the content was actually \textbf{Harmful} or \textbf{Benign}.
\end{enumerate}

The payoff rules are:
\begin{itemize}
    \item If \textbf{Harmful} content is removed, every participant in the group receives \textbf{100 points}.
    \item If \textbf{Benign} content is removed, every participant in the group receives \textbf{0 points}.
    \item \textbf{Remember:} Everyone who chooses to \textbf{Report} pays a cost of \textbf{30 points}, regardless of whether the content is \textbf{Harmful} or \textbf{Benign}, and regardless of whether it is removed.
\end{itemize}

\textbf{In summary, your payoff in each round is shown in the table below (in points).}

\begin{table}[!htbp]
\centering
\begin{tabular}{lcccc}
\toprule
 & \multicolumn{2}{c}{\textbf{Harmful Content}} &
\multicolumn{2}{c}{\textbf{Benign Content}} \\
\cmidrule(lr){2-3} \cmidrule(lr){4-5}
 & Removed & Stays & Removed & Stays \\
\midrule
You Reported & $100 - 30 = 70$ & $-30$ & $-30$ & $-30$ \\[8pt]
You Ignored  & $100$           & $0$   & $0$   & $0$   \\
\bottomrule
\end{tabular}
\end{table}

\paragraph{Bonus Questions}

Before you choose to \textbf{Report}/\textbf{Ignore}, you will also answer two questions. We will describe the questions you need to answer and how we will pay you based on your answers to the questions. If you find the details hard to follow, all you have to remember is that we will pay you in a manner that guarantees that \textbf{it is always in your best interest to report your best guess of the chance that the relevant event happens.}

\textbf{Question 1}

\begin{center}
    \textit{How many of the other 11 participants in your group do you think will choose to \textbf{Report}?}
\end{center}

You will receive \textbf{CNY 10} if your guess is exactly correct; \textbf{CNY 5} if your guess differs from the true number by exactly 1; \textbf{CNY 0} otherwise.

\textbf{Question 2}

\begin{center}
    \textit{What is the probability (0--100\%) that exactly $T-1$ other group members choose to \textbf{Report}?}
\end{center}

Here is exactly how we will pay you for this question: You will submit a guess $X$ for the probability that exactly $T-1$ other group members choose to \textbf{Report}. After you submit your guess of $X$, the interface will draw a value from 0 to 100, with each value being equally likely. Call this value $Y$. The values of $X$, $Y$, and whether or not the event occurs will determine your chances of winning \textbf{CNY 10} or \textbf{CNY 0}.

If $Y$ is greater than or equal to $X$, you will win \textbf{CNY 10} with $Y$\% chance. If $Y$ is less than $X$, you will win \textbf{CNY 10} if the event occurs (if exactly $T-1$ other group members choose to \textbf{Report}).

\textbf{The important thing to remember is that given this payment scheme, it is always in your best interest to choose $X$ that represents your best guess of the chance that the relevant event happens.}

\clearpage

\subsection{Comprehension Quiz}\label{app:quiz}

Participants must answer all quiz questions correctly before proceeding. Incorrect answers trigger an additional review of the instructions, after which the participant answers the quiz again.

\begin{enumerate}
    \item If you see the signal \textsc{LIKELY HARMFUL}, what is the probability that the content is actually harmful?
    \begin{enumerate}[label=(\alph*)]
        \item 50\%
        \item 75\%
        \item 100\% $\checkmark$
        \item Cannot be determined
    \end{enumerate}

    \item Suppose the content is actually \textbf{Benign}. What signal can you receive?
    \begin{enumerate}[label=(\alph*)]
        \item \textsc{Likely Harmful} only
        \item \textsc{Likely Benign} only $\checkmark$
        \item Either \textsc{Likely Harmful} or \textsc{Likely Benign}
        \item Cannot be determined
    \end{enumerate}

    \item Suppose you see the signal \textsc{Likely Benign}. Does this mean the content is definitely benign?
    \begin{itemize}
        \item YES
        \item NO $\checkmark$
    \end{itemize}

    \item If you choose \textbf{Report}, how many points do you pay as a reporting cost?
    \begin{itemize}
        \item 0
        \item 10
        \item 30 $\checkmark$
        \item 100
    \end{itemize}

    \item If you choose \textbf{Ignore}, the content is \textbf{not removed}, and the content is actually \textbf{harmful}, how many points do you earn in that round?
    \begin{itemize}
        \item 100
        \item 70
        \item 30
        \item 0 $\checkmark$
    \end{itemize}

    \item If you choose \textbf{Report}, the content is \textbf{removed}, and the content is actually \textbf{harmful}, how many points do you earn in that round?
    \begin{itemize}
        \item 100
        \item 70 $\checkmark$
        \item 30
        \item -30
    \end{itemize}

    \item Which statement is correct?
    \begin{itemize}
        \item Your group changes from round to round. $\checkmark$
        \item You stay with exactly the same group in every round.
        \item Your decision is revealed to the other participants by name.
        \item The true content type is known before you choose.
    \end{itemize}

    \item \textit{Treatment-specific threshold question. Participants answer only the version assigned to their session.}

    \textit{[High-threshold version:]} If the threshold for removal is $T=9$, when is the content removed?
    \begin{itemize}
        \item When at least 4 participants report
        \item When exactly 9 participants report
        \item When at least 9 participants report $\checkmark$
        \item When all 12 participants report
    \end{itemize}

    \textit{[Medium-threshold version:]} If the threshold for removal is $T=6$, when is the content removed?
    \begin{itemize}
        \item When at least 4 participants report
        \item When exactly 6 participants report
        \item When at least 6 participants report $\checkmark$
        \item When all 12 participants report
    \end{itemize}

    \textit{[Low-threshold version:]} If the threshold for removal is $T=4$, when is the content removed?
    \begin{itemize}
        \item When at least 4 participants report $\checkmark$
        \item When exactly 4 participants report
        \item When at least 9 participants report
        \item When all 12 participants report
    \end{itemize}

    \item What does a higher $\rho$ mean?
    \begin{itemize}
        \item Your signal is more likely to be \textsc{Likely Harmful}
        \item The content is more likely to be harmful
        \item Participants are more likely to receive the same algorithmic report $\checkmark$
        \item The threshold is more likely to be high
    \end{itemize}
\end{enumerate}

Congrats on passing the comprehension quiz! You will now complete 3 practice rounds. These practice rounds are for learning the interface only. Your decisions in the practice rounds do not affect your payment.

Please use these rounds to make sure you understand the timing, the payoff rules, and the meaning of the threshold and the evaluation-report generation process.

\clearpage

\subsection{End Survey}\label{app:end_survey}

After the main experiment, subjects complete the following end survey. Parts A--C are incentivized games. Parts D--I are survey questions and do not directly affect payment.

\subsection*{A. Risk Attitudes (Incentivized Game)}

There is a bag that contains 50 cards. Each card is marked with either the letter R or the letter G. There are 25 R cards and 25 G cards. In each row below, subjects choose between drawing a card from the bag (Choice A) and receiving a corresponding amount for sure (Choice B). If Choice A is implemented, the subject receives CNY 8 if an R card is drawn and receives CNY 0 otherwise.

For payment, one row is randomly selected. The subject's choice in that row is implemented.

\begin{table}[!htbp]
\centering
\begin{tabular}{clcc}
\toprule
Row & Decision & Choice A & Choice B \\
\midrule
1 & Draw a card or receive CNY 0.8 for sure & Draw a card & CNY 0.8 \\
2 & Draw a card or receive CNY 1.6 for sure & Draw a card & CNY 1.6 \\
3 & Draw a card or receive CNY 2.4 for sure & Draw a card & CNY 2.4 \\
4 & Draw a card or receive CNY 3.2 for sure & Draw a card & CNY 3.2 \\
5 & Draw a card or receive CNY 4.0 for sure & Draw a card & CNY 4.0 \\
6 & Draw a card or receive CNY 4.8 for sure & Draw a card & CNY 4.8 \\
7 & Draw a card or receive CNY 5.6 for sure & Draw a card & CNY 5.6 \\
8 & Draw a card or receive CNY 6.4 for sure & Draw a card & CNY 6.4 \\
9 & Draw a card or receive CNY 7.2 for sure & Draw a card & CNY 7.2 \\
\bottomrule
\end{tabular}
\end{table}

\subsection*{B. Altruism / Prosociality (Incentivized Game)}

In this game, each subject is randomly matched with another participant as a pair. In each pair, one participant is assigned the role of Player 1 and the other participant is assigned the role of Player 2. Player 1 allocates CNY 8 between themself and Player 2. For example, if Player 1 allocates CNY $x$ to Player 2, Player 1 receives CNY $8-x$ and Player 2 receives CNY $x$.

\begin{quote}
Suppose you are Player 1. How much would you allocate to Player 2? If you are assigned the role of Player 1, your choice will be implemented.

\medskip
I would allocate CNY \underline{\hspace{2cm}} to Player 2.

\textit{Enter a number between 0 and 8, with up to two decimals allowed.}
\end{quote}

\subsection*{C. Trust (Incentivized Game)}

In this game, two subjects are randomly matched into one pair. One subject is assigned the role of Player A and the other subject is assigned the role of Player B. Player A first chooses between A1 and A2. If Player A chooses A1, Player A receives CNY 4 and Player B receives CNY 0. If Player A chooses A2, Player B chooses between B1 and B2. If Player B chooses B1, both players receive CNY 4. If Player B chooses B2, Player B receives CNY 8 and Player A receives CNY 0.

\begin{center}
\begin{tikzpicture}[
  every node/.style={font=\normalsize},
  edge from parent/.style={draw, -{Stealth[length=6pt]}, thick},
  level 1/.style={sibling distance=35mm, level distance=18mm},
  level 2/.style={sibling distance=32mm, level distance=18mm}
]
\node {A}
  child {node {(4,0)} edge from parent node[left] {A1}}
  child {node {B}
    child {node {(4,4)} edge from parent node[left] {B1}}
    child {node {(0,8)} edge from parent node[right] {B2}}
    edge from parent node[right] {A2}
  };
\end{tikzpicture}
\end{center}

Subjects make choices for the role they may be assigned:

\begin{quote}
Suppose you are Player A. What would you choose if you are assigned the role of Player A?

\medskip
$\circ$ A1 \qquad $\circ$ A2

\medskip
Suppose you are Player B and Player A has chosen A2. What would you choose if you are assigned the role of Player B?

\medskip
$\circ$ B1 \qquad $\circ$ B2
\end{quote}

\subsection*{D. Free-Riding Tendency (Survey)}

\begin{enumerate}
    \item If I believe that enough other participants have already reported the content, I tend to choose Ignore because my own report no longer matters.

    1 = Strongly disagree; 7 = Strongly agree.

    \item In public tasks, if others have already paid the cost, it is morally acceptable for me not to pay the cost while still receiving the benefit.

    1 = Strongly disagree; 7 = Strongly agree.
\end{enumerate}

\subsection*{E. Pivotality Perception (Survey)}

\begin{enumerate}[resume]
    \item When making decisions, to what extent did you consider whether ``my report would be the pivotal vote determining whether the content is removed''?

    0 = Did not consider it at all; 10 = Very important.

    \item When the removal threshold is higher, I feel that my own report is more likely to be useful.

    1 = Strongly disagree; 7 = Strongly agree.

    \item When the removal threshold is lower, I feel that others may already be enough to get the content removed.

    1 = Strongly disagree; 7 = Strongly agree.
\end{enumerate}

\subsection*{F. Understanding of $\rho$ (Survey)}

\begin{enumerate}[resume]
    \item When $\rho$ is higher, which of the following statements is correct?
    \begin{enumerate}[label=(\Alph*)]
        \item My own evaluation report is more accurate.
        \item The content is more likely to be harmful.
        \item Group members are more likely to receive the same evaluation report.
        \item The reporting cost is lower.
    \end{enumerate}

    \item In the experiment, I clearly understood the meaning of $\rho$.

    1 = Strongly disagree; 7 = Strongly agree.
\end{enumerate}

\subsection*{G. Strategic Response Type (Survey)}

\begin{enumerate}[resume]
    \item If I believe that more other group members will report, I myself would be:

    1 = Less willing to report, because others may already be enough.

    4 = Not affected.

    7 = More willing to report, because acting together makes success more likely.
\end{enumerate}

\subsection*{H. Norm / Responsibility (Survey)}

\begin{enumerate}[resume]
    \item Even if my report is unlikely to change the outcome, I feel responsible for reporting as long as the content may be harmful.

    1 = Strongly disagree; 7 = Strongly agree.

    \item If I choose to ignore content that may be harmful, I would feel guilty.

    1 = Strongly disagree; 7 = Strongly agree.
\end{enumerate}

\subsection*{I. Demographics}

\begin{enumerate}[resume]
    \item What is your gender?
    \begin{enumerate}[label=(\Alph*)]
        \item Male
        \item Female
    \end{enumerate}

    \item What is your age?

    Please enter a number.

    \item What is your major or disciplinary background?

    Please enter text.

    \item Have you previously participated in economics experiments?
    \begin{enumerate}[label=(\Alph*)]
        \item Never
        \item 1--2 times
        \item 3 or more times
    \end{enumerate}
\end{enumerate}

\end{document}